\documentclass[twocolumn,aps,pre,superscriptaddress,floatfix]{revtex4-1}
\usepackage{graphicx}
\usepackage{epsfig}
\usepackage{thumbpdf}
\usepackage{amsfonts}
\usepackage{times}     
\usepackage{float}
\usepackage{indentfirst}   %
\usepackage{amsmath}
\usepackage{epstopdf}
\usepackage{textcomp}
\usepackage{tocvsec2}
\usepackage{epsfig}
\usepackage{xcolor}
\usepackage{comment}
\usepackage[normalem]{ulem}
\usepackage{CJK} 
\usepackage[colorlinks=true,linkcolor=black]{hyperref} 
\usepackage{soul}

\newcommand\be{\begin{equation}}
\newcommand\ee{\end{equation}}

\graphicspath{{./Figures/}}

\begin{document}
\begin{CJK}{UTF8}{gbsn}
\title{Effect of network clustering on mutually cooperative coinfections}

\author{Peng-Bi Cui (崔鹏碧)} 
\affiliation{Istituto dei Sistemi Complessi (ISC-CNR), UOS Sapienza, Piazzale A.
Moro 2, 00185 Roma, Italy}
\affiliation{Web Sciences Center, University of Electronic Science and 
Technology of China, Chengdu 611731, China}
\affiliation{Big Data Research Center, University of Electronic Science and
Technology of China, Chengdu 611731, China}

\author{Francesca Colaiori}
\affiliation{Istituto dei Sistemi Complessi (ISC-CNR), UOS Sapienza, Piazzale A.
 Moro 2, 00185 Roma, Italy}
\affiliation{Dipartimento di Fisica, Sapienza Universit\`a di Roma, Roma, Italy}

\author{Claudio Castellano}
\affiliation{Istituto dei Sistemi Complessi (ISC-CNR), Via dei Taurini 19, 00185
 Roma, Italy}

\begin{abstract}
The spread of an infectious disease can be promoted by previous
infections with other pathogens.  This cooperative effect can give
rise to violent outbreaks, reflecting the presence of an abrupt
epidemic transition.  As for other diffusive dynamics, the topology
of the interaction pattern of the host population plays a crucial
role.  It was conjectured that a discontinuous transition arises when
there are relatively few short loops and many long loops in the
contact network.  Here we focus on the role of local clustering in
determining the nature of the transition.  We consider two mutually
cooperative pathogens diffusing in the same population: an individual
already infected with one disease has an increased probability of
getting infected by the other.  We look at how a disease obeying
the susceptible--infected--removed dynamics spreads on contact networks
with tunable clustering.
Using numerical simulations we show that for large cooperativity
the epidemic transition is always abrupt, with the discontinuity 
decreasing as clustering is increased. 
For large clustering strong finite size effects are present and 
the discontinuous nature of the transition is manifest only in large 
networks. 
We also investigate the problem of influential spreaders for
cooperative infections, revealing that both cooperativity and clustering
strongly enhance the dependence of the spreading influence on the
degree of the initial seed.
\end{abstract}

\maketitle
\end{CJK}

\section{Introduction}
\label{sec:introduction}
The modelling of epidemic dynamics is of paramount importance in the
effort to predict when, where, and how far an infectious disease will
spread~\cite{PastorSatorras2015,Kissbook}.  The topological structure
of the social contact network of the host population turns out to play
a key role in determining the patterns of disease transmission.  While
most studies have focused on the dynamics of a single disease,
recently there has been a growing interest in understanding how
concurrent epidemics (coinfections) interact with each other, when
either multiple pathogens or multiple strains of the same disease
simultaneously propagate in the same population.

The interaction among pathogens can have either antagonistic or
synergistic effects.  The main mechanism through which two or more
pathogens spreading in the same population compete is cross--immunity:
An individual infected with one pathogen becomes partially or fully
immune to infection by the others, thus reducing the pool of
susceptible hosts for secondary infections.  The competition between
antagonistic or mutually exclusive epidemics was studied
in~\cite{Newman2005,Funk2010,Marceau2011,Miller2013}.  The opposite
case is the simultaneous spreading of two or more cooperating
pathogens: In this case, an individual already infected with one
disease has increased chance of getting infected by another.  A
notable example is the 1918 ``Spanish" flu pandemic caused by the H1N1
influenza A virus.  The Spanish flu was the the deadliest pandemic in
modern history, involving about one--third of the world's population.
Researchers recently realized that the reason why it was so deadly is
that a considerable proportion of the infected individuals were
coinfected by bacterial pneumonia~\cite{Morens2008,Brundage2008}.
Another well--known example of synergistic effects in disease
spreading is the case of HIV, which increases the host susceptibility
to other pathogens, in particular to the hepatitis C virus
(HCV)~\cite{Sulkowski2008}.

In coinfections, positive feedback between multiple diseases can lead
to sudden and major outbreaks: In 1918 the concurrence of Spanish flu
and pneumonia killed tens of millions of people within
months~\cite{Taubenberger2006}.  One important question in the study
of interacting epidemics is therefore whether cooperation can change
the nature of the epidemic transition from being continuous to being
abrupt when external conditions vary, even slightly, as for a
microscopic change in infectivity.

In Ref.~\cite{Chen2013} a generalized susceptible--infected--removed
(SIR) model (CGCG) was introduced to include {\it mutual} cooperative
effects of co--infections: Two different diseases simultaneously
spread in a population: having been infected with one disease gives an
increased probability to be infected by the other.  The amount of this
increase is a proxy of the mutual {\it cooperativity} between the two
diseases.  The authors studied the model at mean--field level and
observed that cooperative effects, depending of their strength, can
cause a change of the epidemic transition from continuous to
discontinuous.  In~\cite{Janssen2016} Janssen and Stenull showed that
the CGCG model is equivalent, in mean--field, to the homogeneous limit
of an extended general epidemic process~(EGEP)
and clarified the spinodal nature of the discontinuous transition observed.

In Ref.~\cite{Chen2015,Chen2016} the CGCG model was simulated on
lattices and random networks, and it was shown that the type of
transition depends on the contact network topology.  The authors
concluded that a necessary condition for a discontinuous transition to
occur, when starting from a doubly--infected node, is the relative
paucity of short loops with respect to long ones.  A discontinuous
transition occurs if the two epidemics first evolve separately and
then meet only after each of the independent clusters of
singly--infected nodes has reached a large fraction of the population.
At that point, cooperativity between the two pathogens enters into
play, and both clusters rapidly become doubly--infected.  A necessary
condition then is that few short loops are present (otherwise the two
pathogens immediately cooperate and the transition is continuous) and
long loops exist (otherwise cooperativity has no effect and one sees
only single infections).  In agreement with this scenario
discontinuous transitions are absent on trees (no long loops) and on
2--d lattices (many short loops), while they are observed on
Erd{\"o}s--R{\'e}nyi (ER) networks, on 4--d lattices, and on 2--d
lattices with sufficiently long--range contacts\cite{Chen2016}.  In
Ref.~\cite{Cui2017} we have studied the CGCG model on uncorrelated
power--law networks and shown that in the scale--free case, i.e.,
topologies with diverging second moment of the degree distribution,
the transition is always continuous, even for large cooperativity.  On
power--law networks with finite second moment of the degree
distribution, the epidemic transition is instead continuous for low
cooperativity, while it becomes discontinuous when cooperativity is
sufficiently high.  Strong size effects are present, so that the real
nature of the transition is difficult to assess in finite systems.
All the observed discontinuous transitions are of hybrid
type~\cite{Goltsev2006,Parisi2008}: at the transition 
the size of doubly--infected clusters in some realizations
jumps discontinuously from zero to a finite value; 
however the fraction of realizations showing such non--zero clusters
grows continuously from zero at the transition.
(For universal mechanisms underlying hybrid transitions see 
Ref.~\cite{Lee2017}.)

A model for two cooperative infective pathogens not conferring
immunity, analogous to the CGCG model, was recently introduced by Chen
et al.~\cite{Chen2017}.  It is based on the
susceptible--infected--susceptible (SIS) epidemic model and features
increased infectivity if the node susceptible to one pathogen is
already infected
with the other.  By means of numerical simulations on lattices and
homogeneous networks and of a mean--field approach Chen et al. have
shown that in this case large cooperativity can give rise to a
splitting of the epidemic transition in two distinct {\em outbreak}
and {\em eradication} transitions, with associated phenomena of
multistability and hysteresis.  For other recent work about
cooperating infections see 
Refs.~\cite{Sanz2014,HebertDufresne2015,AzimiTafreshi2016}.

Previous work has focused on relatively simple network structures,
where the density of short loops decays to zero as the system size
diverges.  In many real--world topologies instead, in particular those
of social origin, two neighbors of a given node are often mutually
connected, and this property is also observable in very large
networks.  
The clustering coefficient quantifies the abundance of short loops,
by measuring how many of the connected triples form a triangle.
According to the physical argument discussed above, one might
hypothesize that increasing clustering could change the
nature of the epidemic transition from discontinuous to continuous,
even with strong cooperativity. 
In the first part of this paper we focus on the role of
clustering in cooperative epidemics and test in detail this
conjecture.  We generate Poissonian networks with given tunable
clustering using the algorithm introduced by Serrano and
Bogu\~{n}\'a~\cite{Serrano2005}, and study, on the resulting
topologies, the behavior of two cooperating epidemics diffusing
according to the CGCG dynamics.  
A recent work~\cite{Chung2014} has investigated
the same issue, but considering different types of clustered topologies
and of cooperative dynamics.
We find that, in the limit of large networks, 
the epidemic transition is always abrupt and of hybrid nature: 
extensive clusters of nodes hit by both infections suddenly start to
appear at some critical value of the infectivity, 
with a probability that grows from zero at the transition.
The total fraction of nodes belonging to these extensive clusters 
remains finite but becomes smaller when increasing the clustering, 
so that for large clustering it is hard to assess the nature of the 
transition in small systems.  Simulations on large networks, however, 
clearly show a
discontinuous transition also for large clustering.  Our results
indicate that although the paucity of short loops is a necessary
condition to observe the discontinuous epidemic transition, increasing
the density of short loops just by tuning the clustering does not
guarantee to change the nature of the transition to a continuous one.

In the second part of the paper we briefly discuss the problem of
spreading influence for coinfections, i.e., how the probability
that a macroscopic outbreak occurs depends on which node triggers
the coinfection event. We find that large-degree nodes are much
more influential than nodes with few connections than in the
absence of cooperativity. This effect of degree on spreading influence
is further increased when the underlying network is clustered.

\section{The Model for cooperative SIR dynamics} 
\label{sec:model}
Compartmental models are in epidemiology the main mathematical
framework for the study of disease spreading.  In these kind of models
the population is divided in ``compartments" -- in the simplest case
susceptible to the infection (S), infected by the pathogen and able to
transmit it (I), and recovered or removed (immune) (R) -- that
interact according to rules based upon phenomenological assumptions.
Compartmental models branch in two large classes, depending on whether
or not permanent immunity may occur.  Infectious diseases where
recovery confers immunity, such as measles, mumps and rubella are
modelled by susceptible--infected--removed (SIR)~\cite{Kermac1927}
type dynamics: infected individuals transmit the infection to each of
their susceptible neighbors with some probability, while spontaneously
recover with some other probability.  In this case maintaining an
endemic level of infection is impossible in a closed population due to
the depletion of susceptible individuals as the epidemic spreads
through the population.  Other infections, such as the common cold and
some sexually transmitted diseases, do not confer any long lasting
immunity, and after recovery individuals become susceptible again.
These epidemics are modelled by the susceptible--infected--susceptible
(SIS)~\cite{PastorSatorras2015} type of dynamics: the difference with
SIR dynamics is that when infected individuals spontaneously recover
they become again susceptible.  In this paper we only deal with
infections conferring permanent immunity, modeled by SIR dynamics.

In the classical SIR model in discrete time, at each time step each
infected individual spontaneously decays with probability $r$ into the
removed state, while transmitting the infection to each susceptible
neighbor with probability $p$.  The cooperative SIR dynamics that we
consider (CGCG model) is an extension of SIR to two circulating
diseases, A and B as in~\cite{Chen2013}.  The infection probability
for one disease is increased if the individual already contracted the
other disease (even if currently recovered): Individuals uninfected
with either disease get infected (with either A or B) by any infective
neighbor with probability $p$, while a node that is currently infected
or has been infected in the past by with one of the two diseases has a
higher probability $q>p$ to get infected with the other disease by the
neighbor.  When recovering from one disease an individual becomes
immune to it, but can still be infected with the other.  We assume the
same recovery probability $r$ for both diseases, therefore the model
is totally symmetric with respect to A and B.  Since each individual
can be in one of three possible states (S, I, R) with respect to each
of the two diseases (A, B) there are nine possible states for each
individual, denoted as S, A, B, AB, a, b, aB, Ab and ab, where, for
each disease, capital letters refer to the infected state, while
lower--case letters refer to the removed state.  States denoted by
single letters (a, b, A, B) imply that the individual is still
susceptible with respect to the other disease.

\section{Networks with tunable clustering}
\label{sec:algorithm}
In order to perform a detailed analysis of the effects of the topology
on the epidemic dynamics, we generate networks with tunable clustering
to use as contact networks for the epidemics.  To build such networks
we use the algorithm introduced by Serrano and Bogu\~{n}\'a in
Ref.~\cite{Serrano2005}.  The Serrano--Bogu\~{n}\'a algorithm is
devised in the same philosophy of the classical configuration model to
generate maximally random networks of fixed size $N$ with given degree
distribution $P(k)$ {\it and} given average clustering coefficient
$\overline{c}(k)$ for each class of nodes of degree $k$.  Precisely,
$\overline{c}(k)$ denotes the average clustering among all nodes with
degree $k$:
\begin{equation}
\overline{c}(k)=\frac{1}{N_k} \sum^{(k)}_i \frac{2 T_i}
{k(k-1)}
\end{equation}
where $N_k$ is the number of nodes with degree $k$, $T_i$ is the
number of triangles node $i$ belongs to and the sum $\sum^{(k)}$ runs
over all nodes with degree $k$.  The mean clustering coefficient is
then derived by averaging the $\overline{c}(k)$ with the degree
distribution as $\overline{c}=\sum_k P(k) \overline{c}(k)$.

The algorithm is divided in three steps: ({\it i}) assigning a degree
to each node and a number of triangles to each degree class according
to the given distributions; ({\it ii}) closing triangles; ({\it iii})
closing the remaining free stubs as in the classical configuration
model.  More in detail, in the first step, to each vertex is assigned
a degree according to $P(k)$ by attaching to it of a certain number of
stubs (half links).  Also, to each {\it class} of vertices with degree
$k$ is associated a number of triangles according to the given average
clustering coefficient $\overline{c}_k$.  In the second step, stubs
are paired (and eventually unpaired) to form triangles according to
specific rules until each degree class has the desired number of
triangles.  In this step a parameter $\beta_0$ determines the way
stubs are chosen.  This parameter ranges in the interval $ [ 0,1 ]$,
and has the effect of tuning the level of assortativity in the
network: The clustering of high degree nodes is limited by the
presence of degree--correlations in the topology.  As $\beta_0 $
approaches $0$, more assortative networks are produced, which can
accommodate stronger local clustering for nodes with large $k$.  In
the third step of the algorithm the remaining unpaired stubs are
finally closed into links by applying the classical configuration
model algorithm, i.~e. connecting randomly selected pairs of free
stubs, and suspending the restriction on the assigned triangle number.
There are a number of caveats to be considered for the algorithm to
work; for further details we refer to the original paper.

\begin{figure}
\includegraphics[width=\columnwidth]{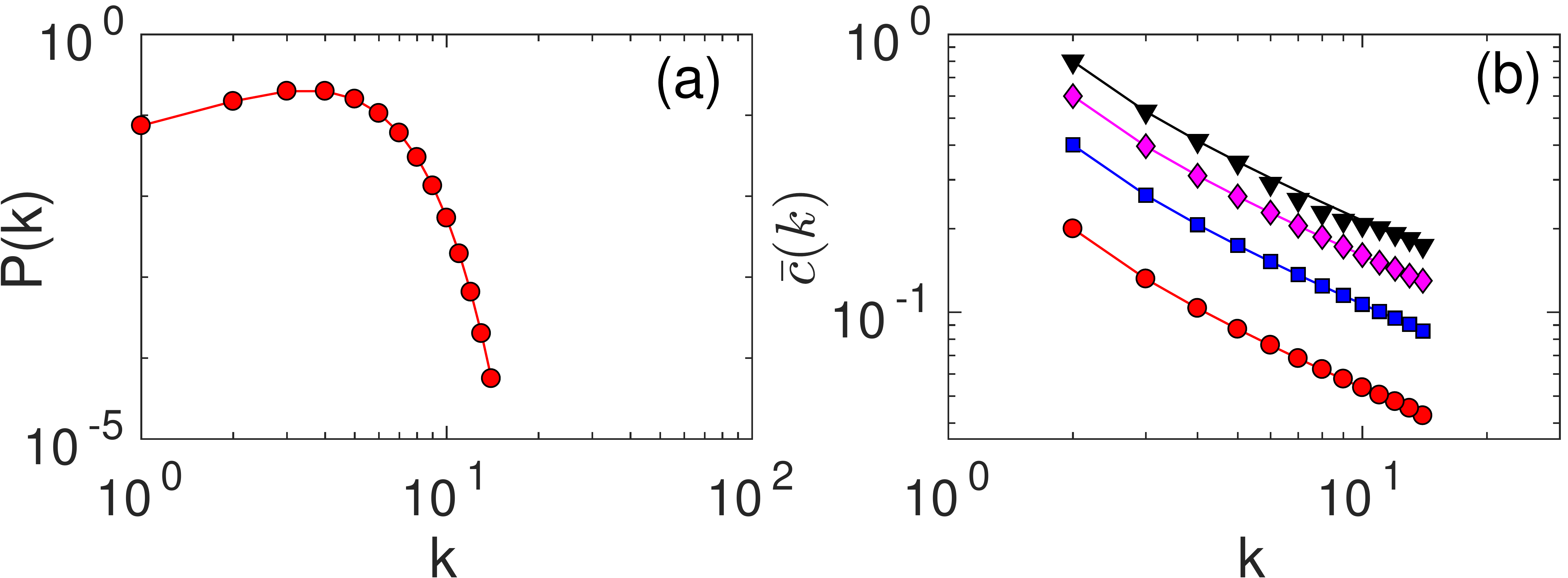}
\caption{(a) Degree distribution generated by the
  Serrano--Bogu\~{n}\'a algorithm using a Poisson degree distribution
  with average degree $z=4$ (red circles). The red solid line represents
  the pre-assigned distribution. (b) Clustering distributions of the
  Poisson networks generated by the Serrano--Bogu\~{n}\'a algorithm
  (symbols). Solid lines represent the pre--assigned clustering
  distributions which are $\overline{c}(k)=c_{0}(k-1)^{-\alpha_{0}}$, with
  $\alpha_{0}=0.6$. From bottom to top, the points represent
  clustering for networks with $c_{0}=0.2$ (red circles), $c_{0}=0.4$
  (blue squares), $c_{0}=0.6$ (pink diamonds), and $c_{0}=0.8$ (black
  triangles), respectively. Data are obtained by averaging over $200$
  realizations. The network size is $N=10^{5}$.
  The parameter $\beta_0$ regulating the assortativity is fixed 
  to $\beta_0=0.1$.}
\label{fig:structure}
\end{figure}

In Fig.~\ref{fig:structure}(a) we show that the distributions
characterizing the networks obtained as outcome of the
Serrano--Bogu\~{n}\'a algorithm closely reproduce the distributions
given as input.  In these examples the algorithm is started with a
Poissonian degree distribution
$P(k)={{N}\choose{k}}(z/N)^k(1-z/N)^{(N-k)}$, where the number or
nodes $N$, and the average degree $z=\langle k\rangle$ are fixed to
$N=10^{5}$, and $z=4$, and with a clustering distribution of the form
$\overline{c}(k)=c_{0}(k-1)^{-\alpha_{0}}$ with $\alpha_{0}$ fixed to
$\alpha_{0}=0.6$ and varying $c_0$.  The value of the parameter
$\beta_0$, that, as explained in Ref.~\cite{Serrano2005}, tunes the
assortativity, is fixed to $\beta_0=0.1$.  The parameter $c_0$ sets
the overall {\it clustering level} of the network and is the key
parameter in our analysis.  A value of $c_0=0$ means that no
clustering is imposed beyond the one naturally occurring for
Poissonian networks.  In what follows we use networks produced by the
Serrano--Bogu\~{n}\'a algorithm as contact patterns for the
cooperative SIR dynamics.  By tuning the $c_0$ parameter we
investigate the effect of loops and clustering on the epidemic
spreading.  All other parameters are kept fixed at the values of
Fig.~\ref{fig:structure}.

\section{The nature of the epidemic transition}
\label{sec:csir}

We simulate the mutually cooperative SIR dynamics ruled by the
CGCG~\cite{Chen2013,Chen2015} model, as defined in
Sec~\ref{sec:model}, with two pathogens $A$ and $B$.  At each time
step, each singly--infected node representing an individual infected
with either $A$ or $B$ attempts to transmit the pathogen to each of
its neighbors that are susceptible to it.  The transmission is
successful with probability $p$ if the neighbor is healthy (in the S
state), and with probability $q>p$ if the neighbor has already
been infected with the other pathogen, even if it has already recovered.
After attempting the contagion, with probability $r$
the singly--infected node recovers from the disease and goes
into the R state.  In the limiting case $q=p$ the
cooperative effect vanishes and the two pathogens spread independently
from one another.

In a similar way, each node in the AB state, representing a
doubly--infected individual, attempts to transmit both pathogens to
each of its neighbors and succeeds in infecting healthy (S) nodes with
just one disease with probability $p$, and with both diseases
simultaneously with probability $p^2$.  Singly--infected neighbors
instead get doubly--infected with probability $q$.  The
doubly--infected node that has attempted the contagion recovers from
either disease shifting into a singly--infected state with probability
$r$, while it recovers from both diseases simultaneously, shifting into
the ab state with probability $r^2$.
In all simulations, unless otherwise stated, we fix $r=1$ (for both
pathogens) and $q=1$, which is the maximum possible value of the
cooperativity.  Unless explicitly stated, we start the system with all
individuals in the susceptible (S) state, except for one randomly
chosen individual who is in the doubly--infected AB state.  

For small values of $p$ only small outbreaks occur,
reaching a finite number of individuals. As $p$ is increased
above the epidemic transition another type of outbreak appears: 
large outbreaks of size proportional to $N$.  The
probability to have a large outbreak grows continuously from zero at
the transition. For this reason the
fraction $\rho_{ab}$ of doubly--recovered nodes in the final state 
averaged over all realizations is not useful for discriminating among a 
continuous and a hybrid transition, as it necessarily changes
continuously.  In order to
discriminate we study the behavior of the average 
$\langle \rho_{ab} \rangle$, computed only on large outbreaks.
Clearly the distinction between large and small outbreaks is
clear--cut only in the infinite size limit.
We operatively define such realizations as those for which in the
final state $\rho_{ab}>T$, where $T$ will be specified below.
Therefore the order parameter for the transition is the average value 
$\langle \rho_{ab} \rangle$, where the average is computed
only over the fraction $P_{ab}$ of realizations fulfilling the 
condition $\rho_{ab}>T$.
\begin{figure}[ht]
\includegraphics[width=\columnwidth]{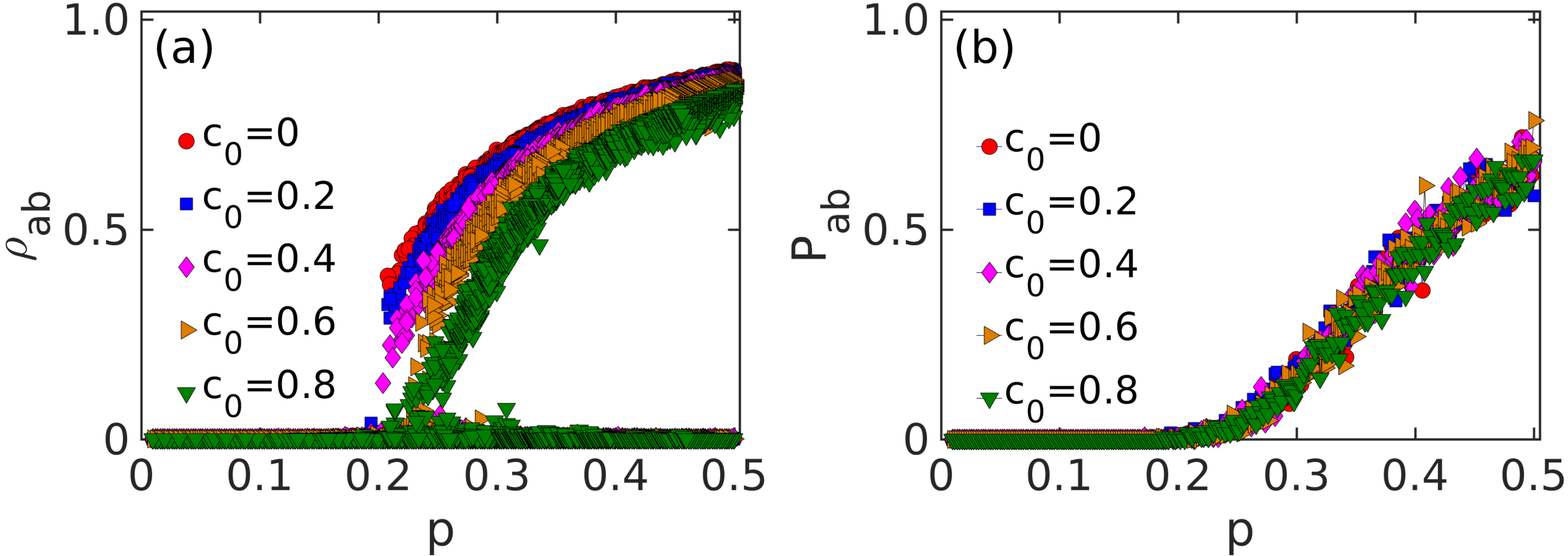}
\caption{
  Results for a single doubly-infected seed.
  (a): 
  Final fraction of population in the doubly--recovered
  state (ab) versus $p$ for networks of size $N=10^4$ and different
  values of $c_0$. Each point is a single realization.  
  There are 200 realizations for each value of $p$. Values of $p$
  are increased by intervals $\Delta p=0.002$.
  (b): Probability $P_{ab}$ that $\rho_{ab}>T=0.005$. 
  Initially all nodes are susceptible except for a randomly
  selected node which is in the doubly--infected state.}
\label{fig:sirclustering}
\end{figure}

In Fig.~\ref{fig:sirclustering} (left) we plot the final fraction of
population in the doubly--recovered state ($\rho_{ab}$) for each
realization and in Fig.~\ref{fig:sirclustering} (right) the probability
$P_{ab}$ that $\rho_{ab}>T$.
When clustering is small, the figure shows a clear discontinuous change:
at a critical value of $p$ some realizations with large $\rho_{ab}$
start to appear. In these runs the two pathogens, starting from the same
doubly--infected node, manage to separately infect large clusters prior
to get in contact with each other. 
At some point, they meet along a large loop and this exposes large 
singly--infected portions of the population to the coinfection that 
diffuses fast given the large coinfectivity, leading to large final
values of $\rho_{ab}$.
When clustering is increased, the jump becomes smaller: 
the abundance of short loops makes it hard for the two epidemics to
develop separated clusters.  
For very high clustering the height of the jump seems to go to zero,
but the nature of the transition cannot be assessed by visual 
inspection: data are inconclusive at this system size.

\begin{figure}
\includegraphics[width=\columnwidth]{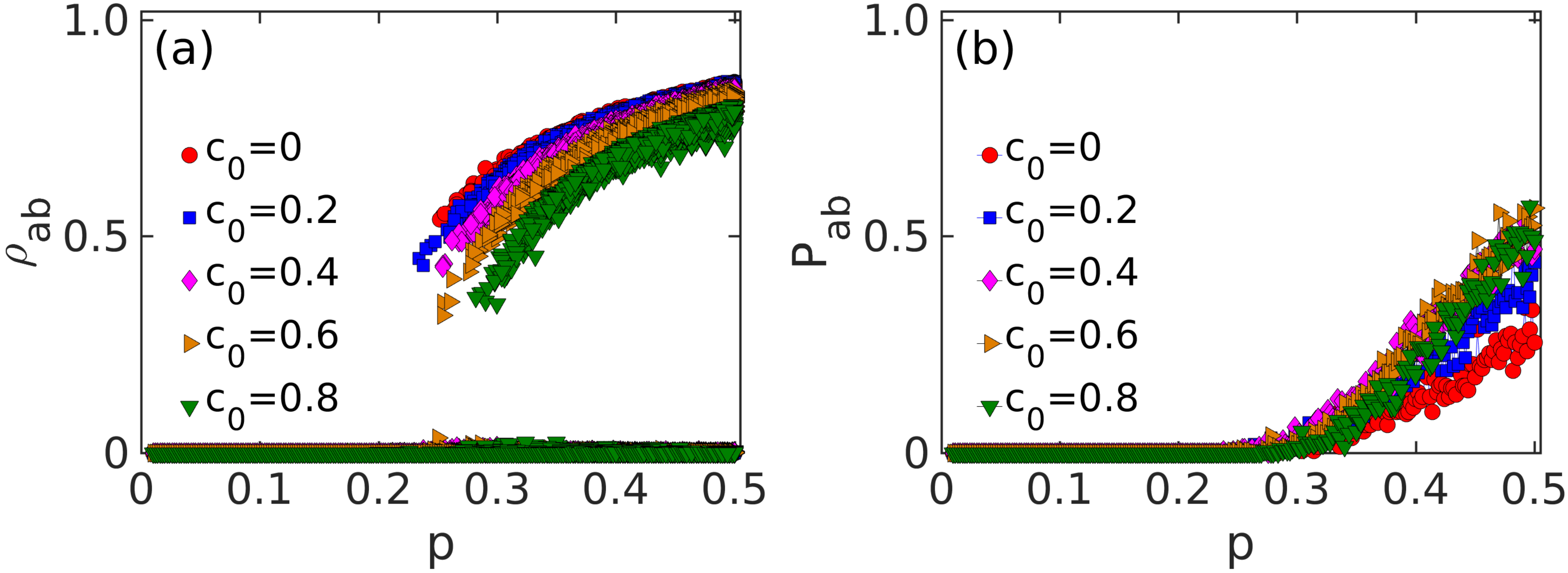}
\caption{
  Results for two distinct singly-infected seeds.
(a): 
  Final fraction of population in the doubly--recovered
  state (ab) versus $p$ for networks of size $N=10^4$ and different
  values of $c_0$. Each point is a single realization.  
  There are 200 realizations for each value of $p$. Values of $p$
  are increased by intervals $\Delta p=0.002$.
  (b): Probability $P_{ab}$ that $\rho_{ab}>T=0.005$. 
  Initially all nodes are susceptible except for two singly--infected
  nodes randomly selected provided the distance between them is larger
  than 8.}
\label{fig:sirsingle}
\end{figure}

To better understand the interpretation in terms of loop structure we
compare the above results with the case where the epidemic is initiated
with two singly--infected nodes, chosen randomly, but constrained
to be at a minimum distance from each other.
In this case for coinfections to occur two large clusters of
singly--infected nodes must necessarily develop first and then meet.  
In this case data show always a large discontinuity for $\rho_{ab}$
and only a weak dependence on the clustering 
(see Fig.~\ref{fig:sirsingle}).
As expected, the presence of short loops does not play any major role 
in this case, in agreement with the physical interpretation in terms 
of loop structure.

\begin{figure}
\includegraphics[width=\columnwidth]{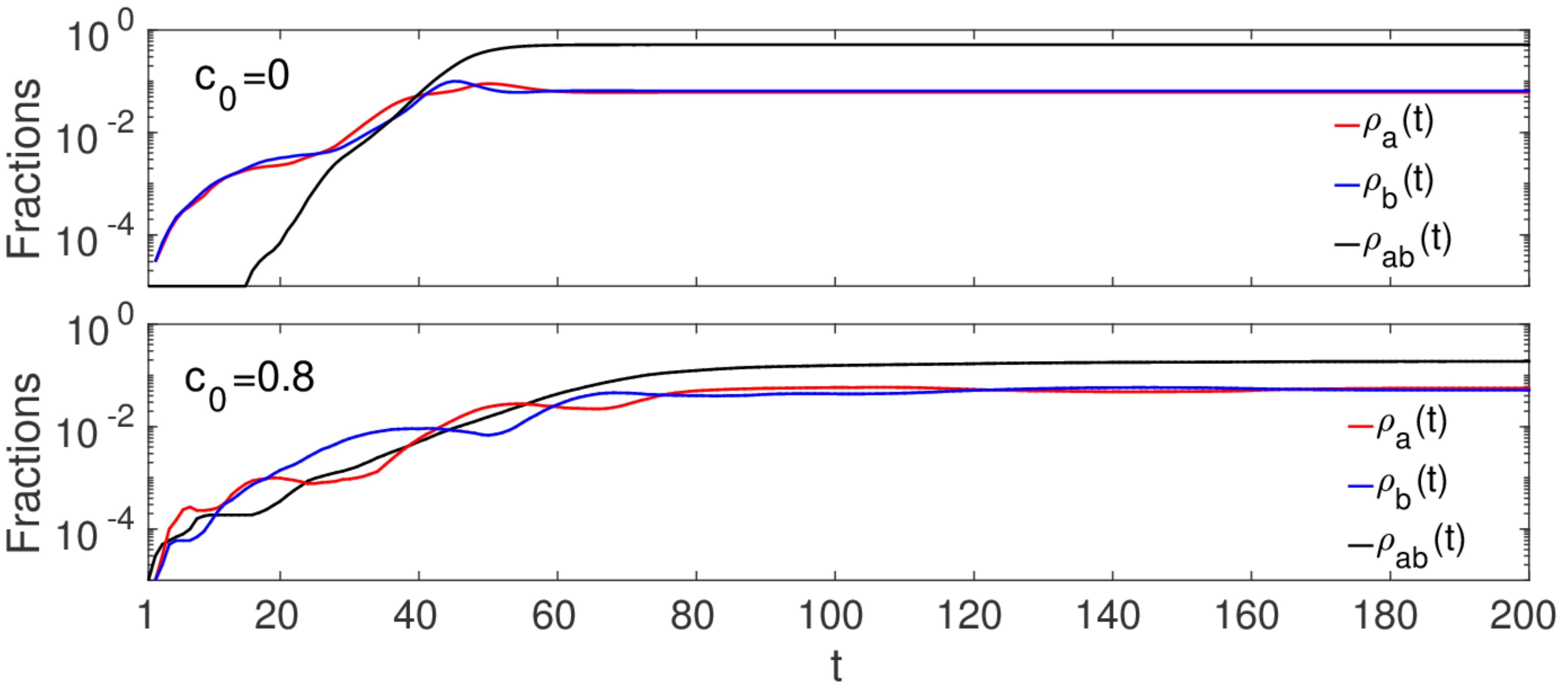}
\caption{Temporal evolution of densities of singly-- or 
  doubly--recovered nodes (states a, b or ab, respectively), 
  doubly-infected nodes (state AB)
  for cooperative SIR dynamics around the threshold ($p=0.25$). The
  network has size $N = 10^5$. Initially all nodes are susceptible
  except for a single randomly-selected node which is in the 
  doubly--infected (AB) state.}
\label{fig:sirtemporal}
\end{figure}

Starting the system with a single doubly-infected seed, we inspected
the temporal evolution of the densities of infected nodes
(Fig.~\ref{fig:sirtemporal}) in two realizations leading to large
epidemics, for a value of $p$ above and around the threshold, in a
system of size $N=10^5$. For no clustering
(Fig.~\ref{fig:sirtemporal}, top) there is an initial transient during
which $\rho_{ab} = 1/N$ (the initial seed) while $\rho_{a}$ and
$\rho_{b}$ rapidly grow, witnessing the formation of large
singly-infected clusters. Around $t = 15$ the two clusters meet over a
long loop and rapidly coinfection takes over.  For large clustering
($c_0 = 0.8$; Fig.~\ref{fig:sirtemporal}, bottom) the dynamics starts
in a way similar to what one would expect in a continuous transition:
A and B infected clusters are intertwined and immediately after being
infected by one pathogen each node is also hit by the other.  As
expected according to the interpretation given in
Refs.~\cite{Chen2015,Chen2016,Cui2017}, the abundance of short loops
causes the two epidemics to meet frequently on the networks and
hampers the independent development of single disease clusters. After
a while, however, doubly infected nodes appear, and the evolution
becomes more similar to the $c_0=0$ case indicating that a similar
scenario, with merging singly-infected clusters, occurs, although on a
smaller scale.  Even by looking closely at the dynamics, the behavior
of the system in the high clustering regime is not clear-cut, and to
identify the nature of the transition a finite size analysis is
needed.

\begin{figure}
\includegraphics[width=0.49\columnwidth]{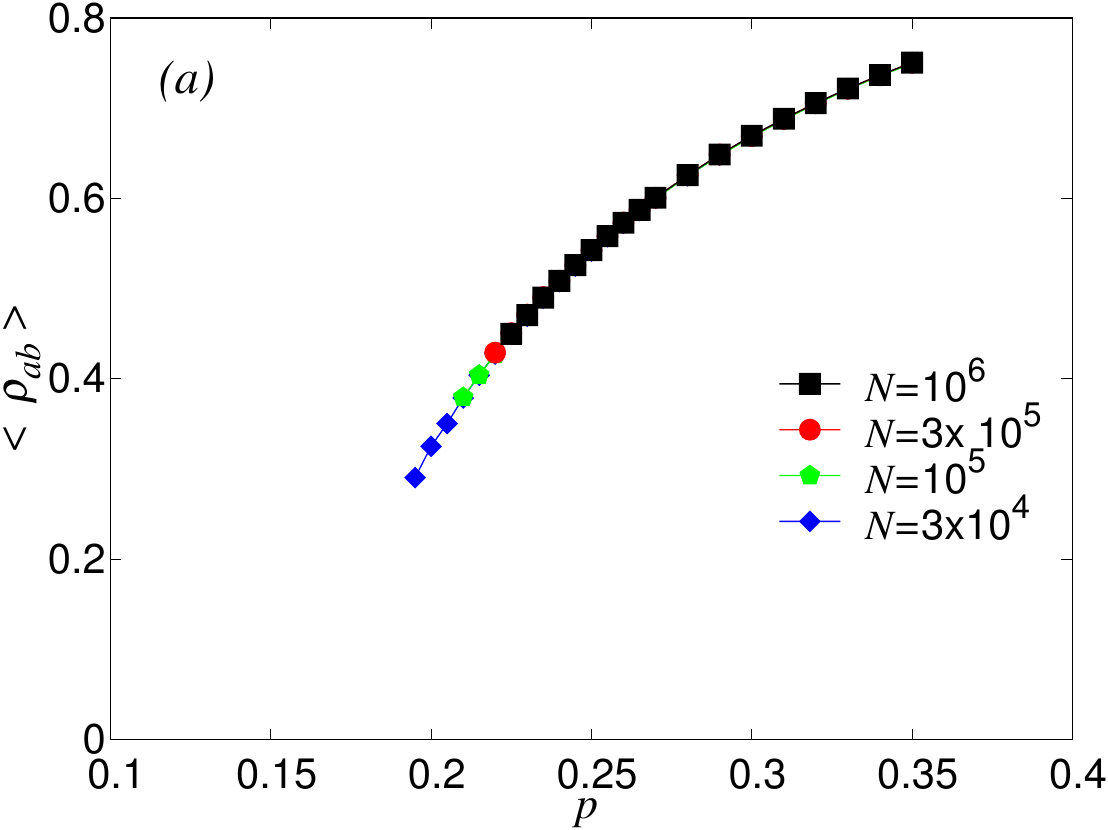}
\includegraphics[width=0.49\columnwidth]{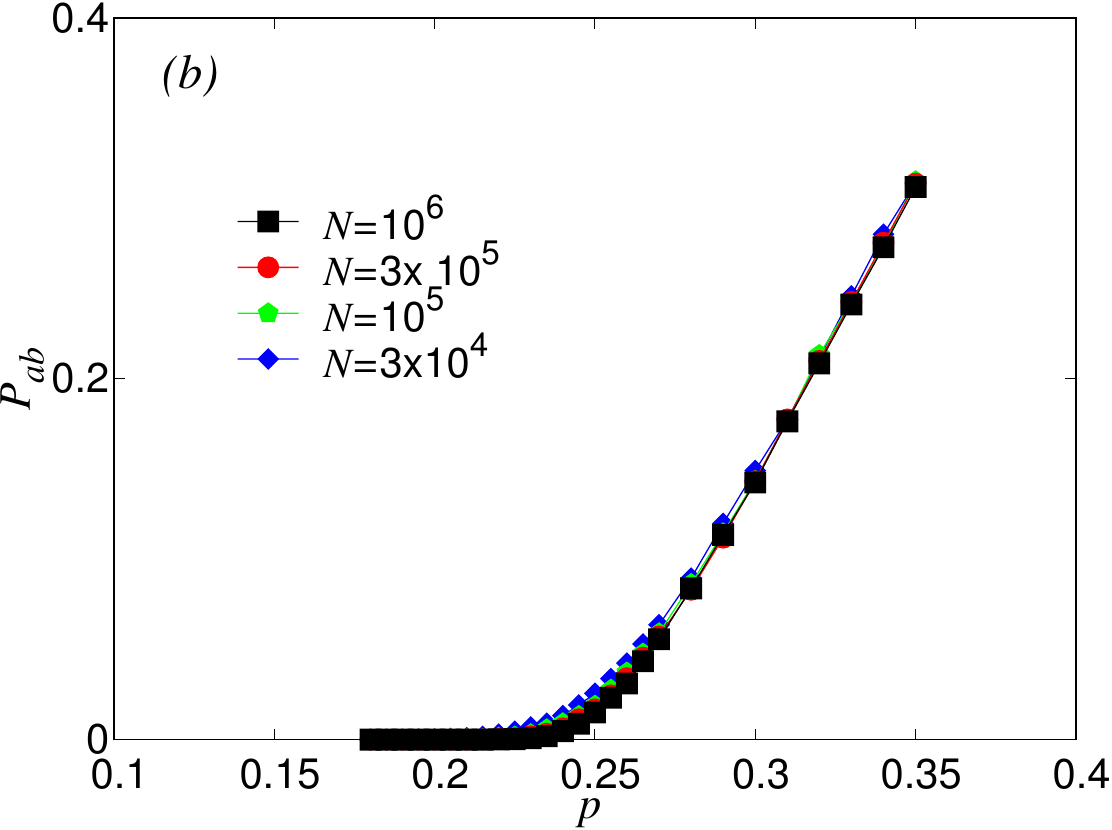}
\includegraphics[width=0.49\columnwidth]{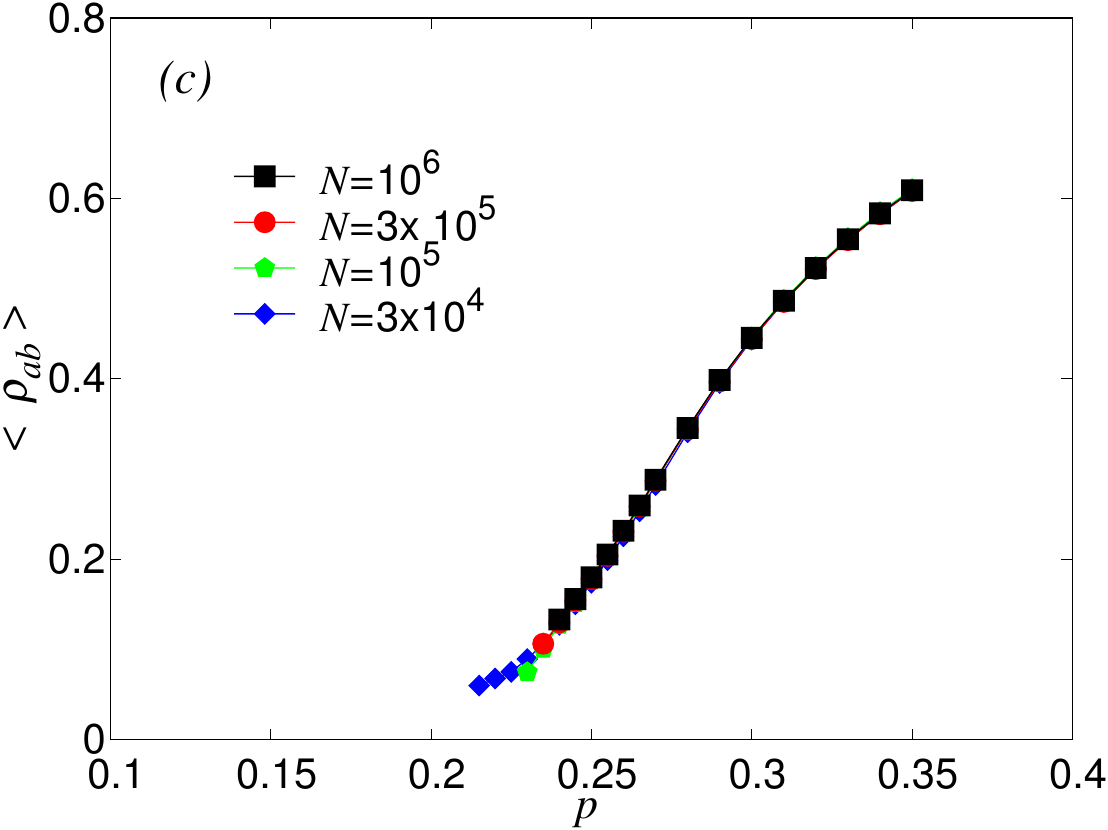}
\includegraphics[width=0.49\columnwidth]{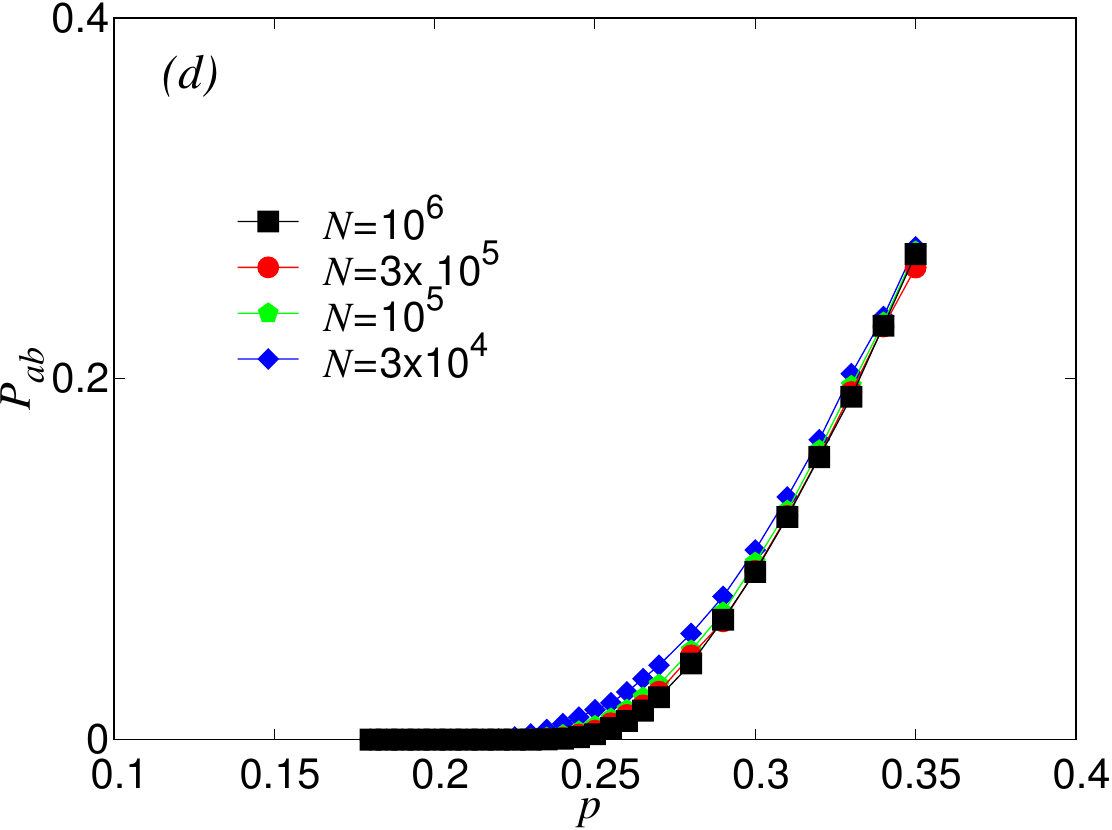}
\caption{Results of simulations for zero ($c_0=0$) and high
  ($c_0=0.8$) connectivity, and different system sizes.  Panels (a)
  and (b) are the results for $c_0=0$ for networks of size $N=3 \times
  10^4, 10^5, 3 \times 10^5,10^6$.  For each value of $N$ the data are
  averaged over $10^5$ realizations: (a) is the average final fraction
  $\langle \rho_{ab} \rangle$ of the population in the 
  doubly--infected state (AB) for large outbreaks versus $p$; (b) is the
  probability $P_{ab}$ that $\rho_{ab}>T=0.05$.  Panels (c) and (d)
  show the same quantities for the case of large clustering,
  $c_0=0.8$. Values of $\langle \rho_{ab} \rangle$ are not plotted
  if the number of realizations to be averaged is less than 10. }
\label{fig:sirc028size}
\end{figure}

As observed in Ref.~\cite{Cui2017} finite size--effects might be
important and conceal the real nature of the transition: To analyze
the effect of network size we simulate the cooperative SIR dynamics on
networks of different sizes, for no clustering ($c_0=0$) and high
clustering $c_0=0.8$.
For each value $p$ of the single disease infectivity parameter we
simulate $N_r=10^5$ realizations of the process starting with a
randomly chosen doubly--infected (AB) seed.  For each realization we
determine the final density $\rho_{ab}$ of nodes in the ab state, and the
probability $P_{ab}$ that $\rho_{ab}>T=0.05$.
After checking the results of $\rho_{ab}$ for single realizations
we fixed the threshold at $T=0.05$, and checked that the results are
unchanged when $T$ is halved.  This ensures that the
$\rho_{ab}$ values are clustered around zero or around some finite
value, and that a sensible gap exists between them.  
We then compute the average value $\langle \rho_{ab} \rangle$ restricted
to values $\rho_{ab} > T$.
Results are shown in Fig.~\ref{fig:sirc028size}. 
In both cases we observe
a hybrid transition: The probability $P_{ab}$ of reaching a finite
value of $\rho_{ab}$ undergoes a continuous transition, for 
$p = p_c \approx 0.25$; on the other hand, the size of the doubly--recovered
(ab) cluster in those realizations jumps discontinuously to a finite value
at the transition.  Large outbreaks can develop in finite systems also
below threshold, however the probability that they occur vanishes as
$N \rightarrow \infty$.
The jump is smaller in the high clustering case, where the discontinuity
is clearly shown only when large enough networks are considered
[Fig.~\ref{fig:sirc028size} (c)].
We conclude that the tendency toward a continuous transition
observed in Fig.~\ref{fig:sirclustering} for large cooperativity
as $c_0$ grows, is a finite-size effect. 
No matter how strong the clustering
the transition is discontinuous in the large size limit. 
The discontinuity arises from the coalescence of 
independently grown singly--infected clusters.
If $p$ is below the threshold value for single epidemics
such extensive singly--infected clusters cannot develop
and coinfectivity does not play any role.
Consistently with this physical interpretation, 
the value of $p$ marking the threshold for coinfections 
coincides with the threshold for single pathogen epidemics.
This threshold slightly increases with the level of clustering $c_0$.

\begin{figure}
\includegraphics[width=0.49\columnwidth]{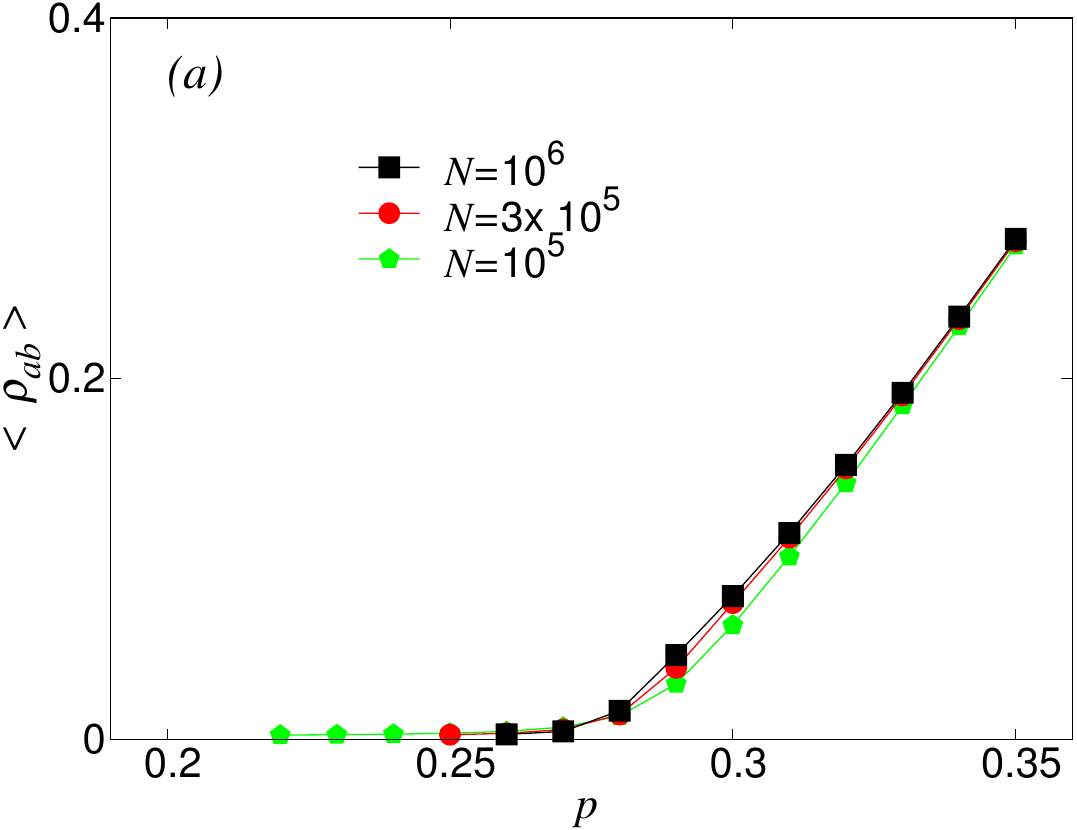}
\includegraphics[width=0.49\columnwidth]{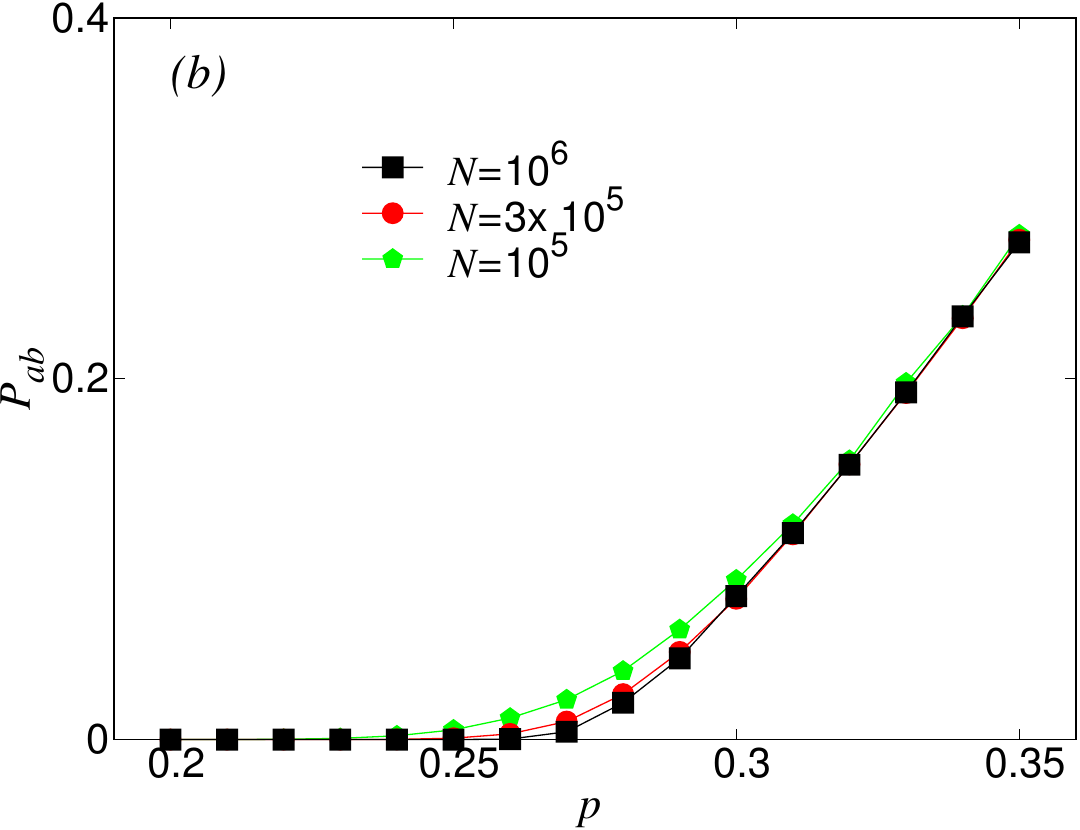}
\caption{ Results for $c=0.8$ and $q=0.4$ for networks of size $N=
  10^5, 3 \times 10^5,10^6$.  For each value of $N$ the data are
  averaged over $10^5$ realizations.  (a) Average final fraction
  $\langle \rho_{ab} \rangle$ in the doubly--recovered state (ab) 
  for large outbreaks versus $p$; (b) Probability $P_{ab}$ that
  $\rho_{ab}>T=0.002$.
  Values of $\langle \rho_{ab} \rangle$ are
  not plotted if the number of realizations with $\rho_{ab}>T$ is less
  than 10.}
\label{q=4}
\end{figure}

For comparison, we repeat a similar analysis for the case of weak
cooperativity ($q=0.4$) and large clustering ($c_0=0.8$).  In this
case the transition remains continuous also in the large size limit,
as shown in Fig.~\ref{q=4}. 
For each system size and for each value
of $p$ the results are averaged over $10^5$ realizations.

\begin{figure}
\includegraphics[width=\columnwidth]{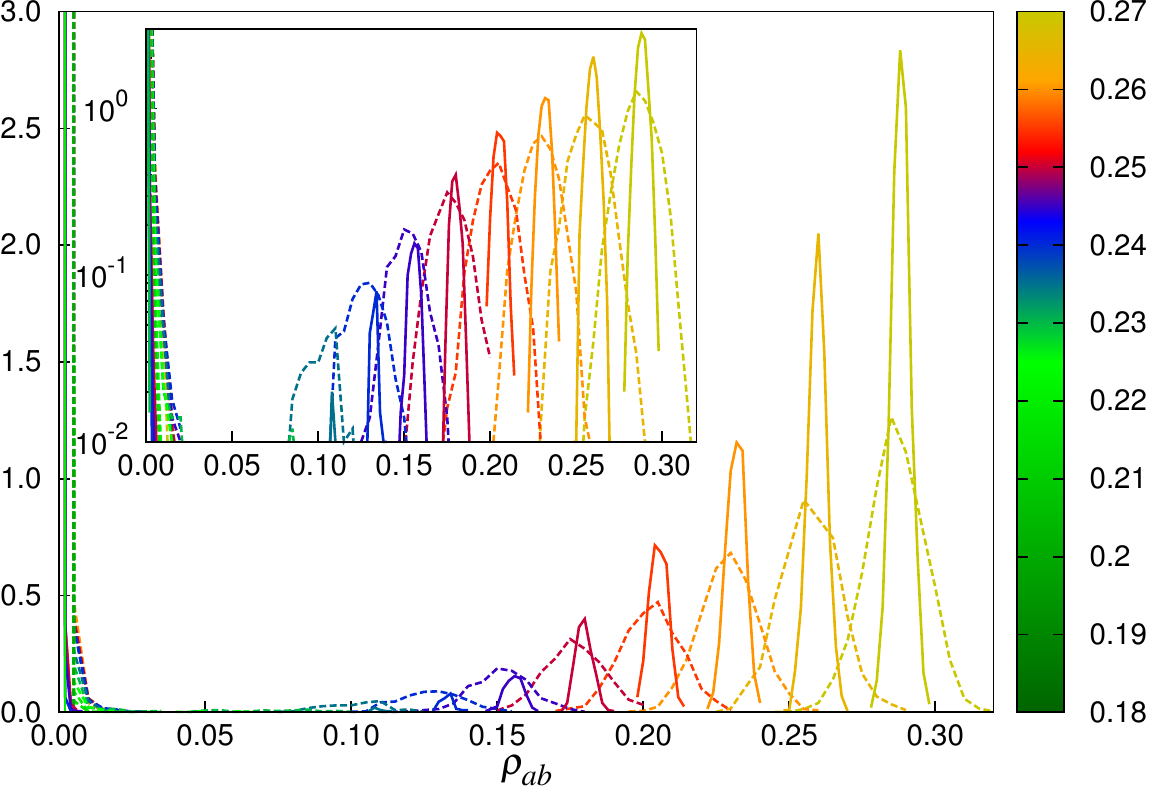}
\includegraphics[width=\columnwidth]{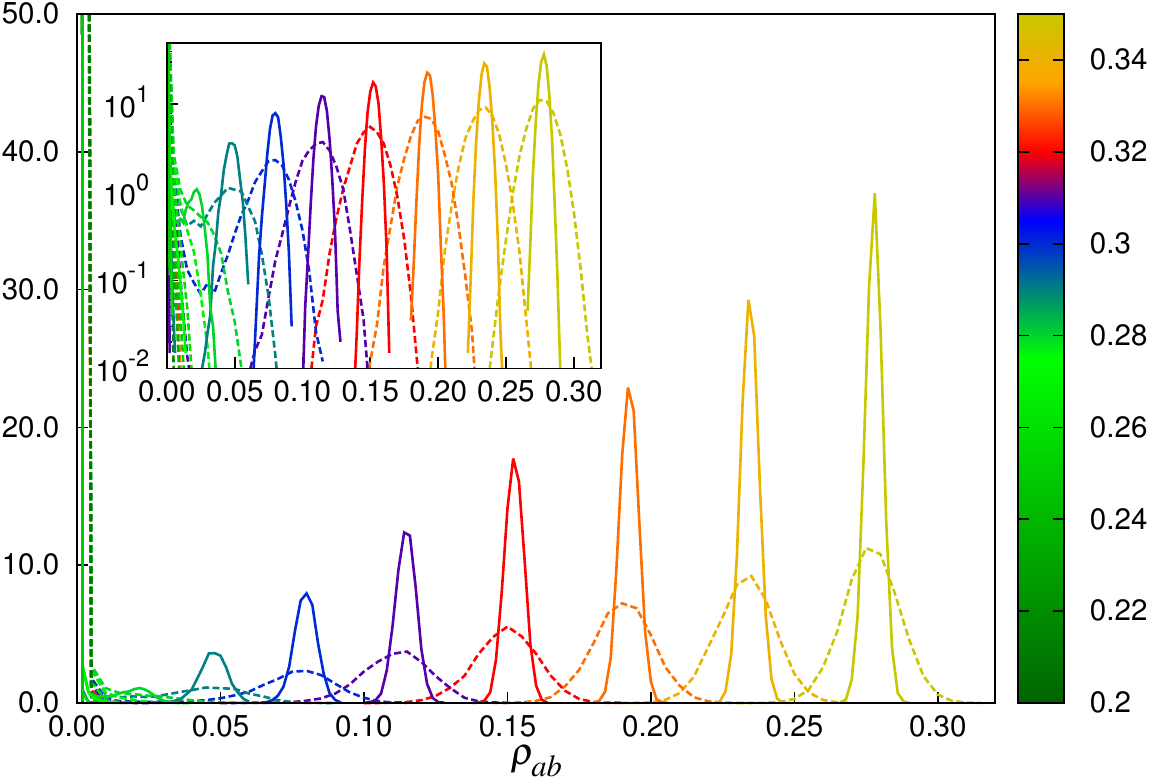}
\caption{Histograms of $\rho_{ab}$ in the doubly--recovered state (ab)
for $N=10^5$ (dotted lines) and $N=10^6$ (solid lines)
and $N_r=10^5$ realizations. Top: $c=0.8$ 
  and $q=1.0$. Each peak corresponds to a value of $p$ starting with
  $p=0.27$ for the rightmost one and decreasing of $\Delta p=0.005$ at
  each curve. The main is in lin--lin scale, the inset is the same plot 
  in lin--log scale. 
  Sharper peaks on the right correspond to the larger system, while 
  on the left they are suppressed as the system size increases, 
  indicating a critical value of $p$ around $0.25$.
  Bottom: $c=0.8$ and $q=0.4$. Each peak
  corresponds to a value of $p$ starting from right with $p=0.35$ and
  decreasing of $\Delta p=0.01$ at each curve. 
  The main is in lin--lin scale, the inset is the same plot in lin--log 
  scale. Sharper peaks correspond to the larger system.  In this case, 
  peaks corresponding to the larger system are never suppressed, consistently
  with a continuous transition.}
\label{hysto}
\end{figure}

Further evidence on the nature of the transition is given by
inspecting the distributions of $\rho_{ab}$ at fixed $p$.  In the case of
high infectivity, even when clustering is high the distribution
shows a secondary peak around a finite value of $\rho_{ab}$ that is
suppressed when the system size grows when $p$ is below a threshold
value around $0.25$.
Fig.~\ref{hysto}, top panel, shows the distributions for the case
$c=0.8$ and two system sizes, $N=10^5$ and $N=10^6$, for $N_r=10^5$
realizations.  Each peak corresponds to a value of $p$ starting with
$p=0.27$ for the rightmost one and decreasing of 
$\Delta p=0.005$ at each curve.
The position of the peaks does not depend on the system
size, as also shown in Fig.~\ref{fig:sirc028size}, where the curves for
different $N$ perfectly overlap.  Fluctuations decrease with the system
size and sharper peaks correspond to the larger systems.
The same data are plotted in lin--log scale in the
inset of the top panel of Fig.~\ref{hysto}:
above some critical value of $p$ (around $0.25$) the height of the
peaks grows for larger $N$; for $p<0.25$ the peaks tend to disappear
as $N$ is increased. This is the signature of a discontinuous transition.
The bottom panel of Fig.~\ref{hysto} shows the same kind of plots, always
for $c_0=0.8$, but for small coinfectivity, $q=0.4$.  Each peak
corresponds to a value of $p$ starting from the right with $p=0.35$ and
decreasing of $\Delta p=0.01$ at each curve.
In this case peaks corresponding to the larger system are never 
suppressed, consistently with the occurrence of a continuous transition.

In a recent work by Chung et al.~\cite{Chung2014},
a model similar to the CGCG, the EGEP,
is found to have an hybrid transition for large cooperativity
in clustered systems.
Our results globally go along the same lines.
However, for some particular values of the parameters determining
the topology, in that work the transition remains continuous 
even for the highest possible cooperativity level, at odds with our results.
This discrepancy may be due to differences in the epidemic dynamics,
since, as pointed out in Ref.~\cite{Janssen2016}, the mapping of the
CGCG model onto the EGEP model only holds within mean--field.

\section{Influential coinfection spreaders}

In this Section, we investigate the problem of identifying
influential spreaders for coinfections. For single spreading 
processes this issue has attracted a lot of interest in
recent years~\cite{Kitsak2010,Lu2016}. 
The problem is the following. The values of $\rho(p)$, the average 
outbreak size, generally considered to study the phase-diagram,
are obtained by averaging over many outbreaks, each starting
in a randomly selected seed. However this quantity is likely to
depend to some extent on the precise location where the infection is seeded.
For example, it is reasonable to expect that nodes with many neighbors
will typically originate larger outbreaks.
Is it possible to predict the spreading influence of node $i$, i.e., 
the average size of outbreaks generated by it?
Is it possible at least to identify topological properties of individual
nodes that are correlated with their spreading influence?

It is clear that degree is positively correlated with $\rho(p)$, but
the detailed structure of the contact pattern makes in some cases 
centralities such as the $k$-core index, betweenness or eigenvalue 
or other centralities,  better predictors of the spreading 
influence~\cite{Lu2016}.
The mapping of SIR dynamics to bond 
percolation~\cite{Grassberger1983,Newman2002,Karrer2014} allows,
at the epidemic threshold $p=p_c$, to identify the Non-Backtracking 
centrality~\cite{Martin2014} as the 
exact solution (i.e. a centrality perfectly correlated with the spreading 
influence) on locally tree-like networks~\cite{Radicchi2016}.
On the same type of topologies, the spreading influence of each individual
node can be exactly calculated for any value of $p$
by message-passing techniques~\cite{Min2018}.

For cooperating epidemics, the problem is slightly different.
As the transition is discontinuous, the most interesting observable 
is the probability $P_{ab}(i)$ that seeding the double-infection in node
$i$ will generate a macroscopic double-infection outbreak.

\begin{figure}
\includegraphics[width=\columnwidth]{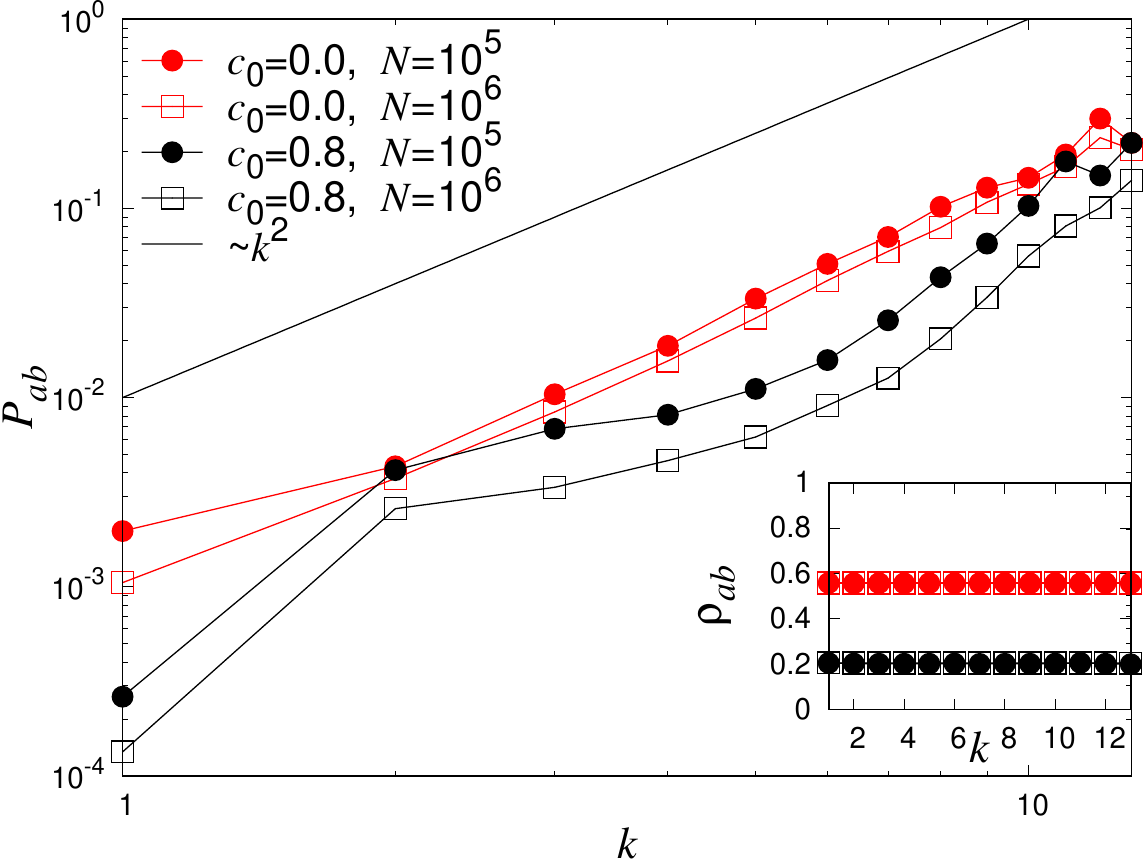}
\caption{
Results from simulations on networks with $c_0=0$ (red) and
$c_0=0.8$ (black), with $p=0.255$. 
System size is $N=10^5$ (filled symbols) and $N=10^6$
(empty symbols).
Main: probability $P_{ab}$ that a double-infection 
started in a node of degree $k$ originates a macroscopic outbreak.
Inset: Average relative size of the macroscopic outbreaks when the infection
is seeded in a node of degree $k$.}
\label{fig:influence}
\end{figure}

We have simulated the coinfection process with initial condition given by 
a doubly-infected seed in each node of the network, for a value of
$p$ immediately above the epidemic transition.
We have computed the fraction of times an outbreak of relative
size larger than $T=0.15$ is produced 
and also the average size of these macroscopic outbreaks. 
We have then averaged the results over nodes with the same degree $k$
(see Fig.~\ref{fig:influence}). 
In the absence of clustering the probability to give rise to a macroscopic 
outbreak is proportional to the square of the node degree. 
In the presence of clustering instead the probability $P_{ab}$ is overall 
smaller, but its growth with $k$ is much faster.
In both cases the size of the extensive coinfection is, with remarkable
accuracy, independent from $k$.

This behavior can be rationalized in the light of our understanding of
the physical origin of large outbreaks for cooperating infections.
Let us first consider unclustered topologies.
For a large outbreak to occur one needs the separate development
of two single epidemics along different directions. The size
of each of them is on average proportional to the degree $k$ 
of the seed node. Hence the probability that they meet is
proportional to $k^2$.
Once the two infections have met, the coinfection rapidly
spreads throughout the network. This last part of the process 
clearly does not depend on the degree of the initial seed.

In the presence of local clustering it is much more difficult for
the two single epidemics to evolve separately, because of the
presence of triangles. This explains the reduction of $P_{ab}$ with 
respect to the unclustered case. 
However, since the clustering coefficient strongly decreases with $k$ 
(see Fig.~\ref{fig:structure}) the spreading influence of nodes 
grows fast with the number of neighbors.
Also in this case the formation of the macroscopic outbreak
after the two independent epidemics have met does not depend
on local properties of the seed node, i.e. it is independent from $k$.

\section{Conclusions}
\label{sec:conclusions}

In Ref.~\cite{Chen2015} it was shown that the type of transition depends
on the topology of the network.  It was suggested that necessary
conditions to observe a discontinuous transition when starting from a
doubly--infected node are the paucity of short loops coupled with the
existence of long loops.  A discontinuous transition occurs if the two
clusters of singly--infected nodes spread independently, and then
meet, rapidly becoming doubly--infected due to large cooperativity.
The authors concluded that this can happen if few short loops are present
--to allow the growth of two independent single--infection clusters--
and long loops exist --to allow the two clusters to meet.  
Based on this physical picture one can hypothesize that in
topologies with many short loops the two diseases do not have the
possibility to spread in independent regions; from the
beginning they form a single doubly--infected cluster which, 
in a way similar to the single disease case, gives rise to a 
continuous transition as infectivity is increased.  

We have analyzed the effect of local network clustering on the spreading of
two concurrent cooperating diseases under SIR dynamics,
i.e. infections conferring permanent immunity.  A detailed analysis
has been performed by means of numerical simulations on Poissonian
contact networks with tunable clustering. 
Our simulations show that increasing the number of loops of length 3,
i.e., the local clustering, does not produce this effect.
The epidemic transition remains discontinuous, although
the size of the jump is reduced when clustering is increased. 
The nature of the transition is evident only in large networks, 
while for small sizes it is blurred by fluctuations.

The observed behavior suggests the following refined physical picture, 
that builds upon the one proposed in Ref.~\cite{Chen2015} and further
clarifies the role of the network loop structure in determining the nature 
of the transition.  
We hypothesize that it is still possible to
observe an abrupt transition when short loops abound, as long as a gap
exists between short and long loops.  The proposed scenario would
explain the above finite size effects in high clustered networks as
follows.  In large systems with high clustering there are many loops
of length 3, however longer loops but still smaller than the
network diameter, that is of order $\log N$, are scarce. Their number
decays as $1/N$.  Although rare, single epidemics that manage to
escape the structure of loops of length 3 can therefore grow independently
before they meet.  For small networks however, the loop structure is
not so well separated in small and large loops for two reasons: on the 
one hand the diameter is smaller, on the other hand also loops of size
smaller than $\log N$ become more abundant.  In this case, single
epidemics that do not meet along loops of length 3 have some probability
to meet along loops of intermediate length. This makes the distribution of
$\rho_{ab}$ broader, and the discontinuous transition more
difficult to identify.
It would be interesting to investigate the role of the loop structure of
the network beyond the clustering in cooperative coinfections.

Concerning spreading influence for coinfections, our results for unclustered
networks nicely fit with the physical picture outlined above. The $k^2$
dependence of $P_{ab}$ on the seed degree is perfectly consistent with the
scenario of two single infections evolving separately and then meeting.
Also the stronger growth with $k$ in the presence of clustering can be
interpreted along the same lines, as the effect of the reduction of the
local clustering coefficient with the degree of the seed.
In both cases the effect of the degree on the spreading influence of a node
is much stronger than for single epidemics. Hence it is even more crucial in
this case to monitor and possibly immunize hubs in order to prevent
extensive coinfection outbreaks in the system. 
It would be extremely interesting to attack these problems, analyzed here 
by means of numerical simulations, using message-passing techniques.

\section{Acknowledgments}
This work was supported by China Scholarship Council
(CSC), grant number 201606075021.

\bibliography{multireference}

\begin{thebibliography}{33}%
\makeatletter
\providecommand \@ifxundefined [1]{%
 \@ifx{#1\undefined}
}%
\providecommand \@ifnum [1]{%
 \ifnum #1\expandafter \@firstoftwo
 \else \expandafter \@secondoftwo
 \fi
}%
\providecommand \@ifx [1]{%
 \ifx #1\expandafter \@firstoftwo
 \else \expandafter \@secondoftwo
 \fi
}%
\providecommand \natexlab [1]{#1}%
\providecommand \enquote  [1]{``#1''}%
\providecommand \bibnamefont  [1]{#1}%
\providecommand \bibfnamefont [1]{#1}%
\providecommand \citenamefont [1]{#1}%
\providecommand \href@noop [0]{\@secondoftwo}%
\providecommand \href [0]{\begingroup \@sanitize@url \@href}%
\providecommand \@href[1]{\@@startlink{#1}\@@href}%
\providecommand \@@href[1]{\endgroup#1\@@endlink}%
\providecommand \@sanitize@url [0]{\catcode `\\12\catcode `\$12\catcode
  `\&12\catcode `\#12\catcode `\^12\catcode `\_12\catcode `\%12\relax}%
\providecommand \@@startlink[1]{}%
\providecommand \@@endlink[0]{}%
\providecommand \url  [0]{\begingroup\@sanitize@url \@url }%
\providecommand \@url [1]{\endgroup\@href {#1}{\urlprefix }}%
\providecommand \urlprefix  [0]{URL }%
\providecommand \Eprint [0]{\href }%
\providecommand \doibase [0]{http://dx.doi.org/}%
\providecommand \selectlanguage [0]{\@gobble}%
\providecommand \bibinfo  [0]{\@secondoftwo}%
\providecommand \bibfield  [0]{\@secondoftwo}%
\providecommand \translation [1]{[#1]}%
\providecommand \BibitemOpen [0]{}%
\providecommand \bibitemStop [0]{}%
\providecommand \bibitemNoStop [0]{.\EOS\space}%
\providecommand \EOS [0]{\spacefactor3000\relax}%
\providecommand \BibitemShut  [1]{\csname bibitem#1\endcsname}%
\let\auto@bib@innerbib\@empty
\bibitem [{\citenamefont {Pastor-Satorras}\ \emph {et~al.}(2015)\citenamefont
  {Pastor-Satorras}, \citenamefont {Castellano}, \citenamefont {Van~Mieghem},\
  and\ \citenamefont {Vespignani}}]{PastorSatorras2015}%
  \BibitemOpen
  \bibfield  {author} {\bibinfo {author} {\bibfnamefont {R.}~\bibnamefont
  {Pastor-Satorras}}, \bibinfo {author} {\bibfnamefont {C.}~\bibnamefont
  {Castellano}}, \bibinfo {author} {\bibfnamefont {P.}~\bibnamefont
  {Van~Mieghem}}, \ and\ \bibinfo {author} {\bibfnamefont {A.}~\bibnamefont
  {Vespignani}},\ }\href {\doibase 10.1103/RevModPhys.87.925} {\bibfield
  {journal} {\bibinfo  {journal} {Rev. Mod. Phys.}\ }\textbf {\bibinfo {volume}
  {87}},\ \bibinfo {pages} {925} (\bibinfo {year} {2015})}\BibitemShut
  {NoStop}%
\bibitem [{\citenamefont {Kiss}\ \emph {et~al.}(2017)\citenamefont {Kiss},
  \citenamefont {Miller},\ and\ \citenamefont {Simon}}]{Kissbook}%
  \BibitemOpen
  \bibfield  {author} {\bibinfo {author} {\bibfnamefont {I.~Z.}\ \bibnamefont
  {Kiss}}, \bibinfo {author} {\bibfnamefont {J.~C.}\ \bibnamefont {Miller}}, \
  and\ \bibinfo {author} {\bibfnamefont {P.}~\bibnamefont {Simon}},\ }\href
  {http://www.springer.com/gp/book/9783319508047} {\emph {\bibinfo {title}
  {Mathematics of epidemics on networks: from exact to approximate models}}}\
  (\bibinfo  {publisher} {Springer},\ \bibinfo {year} {2017})\BibitemShut
  {NoStop}%
\bibitem [{\citenamefont {Newman}(2005{\natexlab{a}})}]{Newman2005}%
  \BibitemOpen
  \bibfield  {author} {\bibinfo {author} {\bibfnamefont {M.~E.~J.}\
  \bibnamefont {Newman}},\ }\href {\doibase 10.1103/PhysRevLett.95.108701}
  {\bibfield  {journal} {\bibinfo  {journal} {Phys. Rev. Lett.}\ }\textbf
  {\bibinfo {volume} {95}},\ \bibinfo {pages} {108701} (\bibinfo {year}
  {2005}{\natexlab{a}})}\BibitemShut {NoStop}%
\bibitem [{\citenamefont {Funk}\ and\ \citenamefont {Jansen}(2010)}]{Funk2010}%
  \BibitemOpen
  \bibfield  {author} {\bibinfo {author} {\bibfnamefont {S.}~\bibnamefont
  {Funk}}\ and\ \bibinfo {author} {\bibfnamefont {V.~A.~A.}\ \bibnamefont
  {Jansen}},\ }\href {\doibase 10.1103/PhysRevE.81.036118} {\bibfield
  {journal} {\bibinfo  {journal} {Phys. Rev. E}\ }\textbf {\bibinfo {volume}
  {81}},\ \bibinfo {pages} {036118} (\bibinfo {year} {2010})}\BibitemShut
  {NoStop}%
\bibitem [{\citenamefont {Marceau}\ \emph {et~al.}(2011)\citenamefont
  {Marceau}, \citenamefont {No\"el}, \citenamefont {H\'ebert-Dufresne},
  \citenamefont {Allard},\ and\ \citenamefont {Dub\'e}}]{Marceau2011}%
  \BibitemOpen
  \bibfield  {author} {\bibinfo {author} {\bibfnamefont {V.}~\bibnamefont
  {Marceau}}, \bibinfo {author} {\bibfnamefont {P.-A.}\ \bibnamefont {No\"el}},
  \bibinfo {author} {\bibfnamefont {L.}~\bibnamefont {H\'ebert-Dufresne}},
  \bibinfo {author} {\bibfnamefont {A.}~\bibnamefont {Allard}}, \ and\ \bibinfo
  {author} {\bibfnamefont {L.~J.}\ \bibnamefont {Dub\'e}},\ }\href {\doibase
  10.1103/PhysRevE.84.026105} {\bibfield  {journal} {\bibinfo  {journal} {Phys.
  Rev. E}\ }\textbf {\bibinfo {volume} {84}},\ \bibinfo {pages} {026105}
  (\bibinfo {year} {2011})}\BibitemShut {NoStop}%
\bibitem [{\citenamefont {Miller}(2013)}]{Miller2013}%
  \BibitemOpen
  \bibfield  {author} {\bibinfo {author} {\bibfnamefont {J.~C.}\ \bibnamefont
  {Miller}},\ }\href {\doibase 10.1103/PhysRevE.87.060801} {\bibfield
  {journal} {\bibinfo  {journal} {Phys. Rev. E}\ }\textbf {\bibinfo {volume}
  {87}},\ \bibinfo {pages} {060801} (\bibinfo {year} {2013})}\BibitemShut
  {NoStop}%
\bibitem [{\citenamefont {Morens}\ \emph {et~al.}(2008)\citenamefont {Morens},
  \citenamefont {Taubenberger},\ and\ \citenamefont {Fauci}}]{Morens2008}%
  \BibitemOpen
  \bibfield  {author} {\bibinfo {author} {\bibfnamefont {D.~M.}\ \bibnamefont
  {Morens}}, \bibinfo {author} {\bibfnamefont {J.~K.}\ \bibnamefont
  {Taubenberger}}, \ and\ \bibinfo {author} {\bibfnamefont {A.~S.}\
  \bibnamefont {Fauci}},\ }\href {\doibase 10.1086/591708} {\bibfield
  {journal} {\bibinfo  {journal} {The Journal of Infectious Diseases}\ }\textbf
  {\bibinfo {volume} {198}},\ \bibinfo {pages} {962} (\bibinfo {year}
  {2008})}\BibitemShut {NoStop}%
\bibitem [{\citenamefont {Brundage}\ and\ \citenamefont
  {Shanks}(2008)}]{Brundage2008}%
  \BibitemOpen
  \bibfield  {author} {\bibinfo {author} {\bibfnamefont {J.~F.}\ \bibnamefont
  {Brundage}}\ and\ \bibinfo {author} {\bibfnamefont {G.}~\bibnamefont
  {Shanks}},\ }\href {\doibase https://dx.doi.org/10.3201/eid1408.071313}
  {\bibfield  {journal} {\bibinfo  {journal} {Emerg. Infect Dis.}\ }\textbf
  {\bibinfo {volume} {14}},\ \bibinfo {pages} {1193} (\bibinfo {year}
  {2008})}\BibitemShut {NoStop}%
\bibitem [{\citenamefont {Sulkowski}(2008)}]{Sulkowski2008}%
  \BibitemOpen
  \bibfield  {author} {\bibinfo {author} {\bibfnamefont {M.~S.}\ \bibnamefont
  {Sulkowski}},\ }\href {\doibase http://dx.doi.org/10.1016/j.jhep.2007.11.009}
  {\bibfield  {journal} {\bibinfo  {journal} {Journal of Hepatology}\ }\textbf
  {\bibinfo {volume} {48}},\ \bibinfo {pages} {353} (\bibinfo {year}
  {2008})}\BibitemShut {NoStop}%
\bibitem [{\citenamefont {Taubenberger}\ and\ \citenamefont
  {Morens}(2006)}]{Taubenberger2006}%
  \BibitemOpen
  \bibfield  {author} {\bibinfo {author} {\bibfnamefont {J.~K.}\ \bibnamefont
  {Taubenberger}}\ and\ \bibinfo {author} {\bibfnamefont {D.~M.}\ \bibnamefont
  {Morens}},\ }\href {\doibase 10.3201/eid1201.050979} {\bibfield  {journal}
  {\bibinfo  {journal} {Emerging Infectious Diseases}\ }\textbf {\bibinfo
  {volume} {12}},\ \bibinfo {pages} {15} (\bibinfo {year} {2006})}\BibitemShut
  {NoStop}%
\bibitem [{\citenamefont {Chen}\ \emph {et~al.}(2013)\citenamefont {Chen},
  \citenamefont {Ghanbarnejad}, \citenamefont {Cai},\ and\ \citenamefont
  {Grassberger}}]{Chen2013}%
  \BibitemOpen
  \bibfield  {author} {\bibinfo {author} {\bibfnamefont {L.}~\bibnamefont
  {Chen}}, \bibinfo {author} {\bibfnamefont {F.}~\bibnamefont {Ghanbarnejad}},
  \bibinfo {author} {\bibfnamefont {W.}~\bibnamefont {Cai}}, \ and\ \bibinfo
  {author} {\bibfnamefont {P.}~\bibnamefont {Grassberger}},\ }\href {\doibase
  10.1209/0295-5075/104/50001} {\bibfield  {journal} {\bibinfo  {journal} {EPL
  (Europhysics Letters)}\ }\textbf {\bibinfo {volume} {104}},\ \bibinfo {pages}
  {50001} (\bibinfo {year} {2013})}\BibitemShut {NoStop}%
\bibitem [{\citenamefont {Janssen}\ and\ \citenamefont
  {Stenull}(2016)}]{Janssen2016}%
  \BibitemOpen
  \bibfield  {author} {\bibinfo {author} {\bibfnamefont {H.-K.}\ \bibnamefont
  {Janssen}}\ and\ \bibinfo {author} {\bibfnamefont {O.}~\bibnamefont
  {Stenull}},\ }\href {\doibase 10.1209/0295-5075/113/26005} {\bibfield
  {journal} {\bibinfo  {journal} {EPL (Europhysics Letters)}\ }\textbf
  {\bibinfo {volume} {113}},\ \bibinfo {pages} {26005} (\bibinfo {year}
  {2016})}\BibitemShut {NoStop}%
\bibitem [{\citenamefont {Cai}\ \emph {et~al.}(2015)\citenamefont {Cai},
  \citenamefont {Chen}, \citenamefont {Ghanbarnejad},\ and\ \citenamefont
  {Grassberger}}]{Chen2015}%
  \BibitemOpen
  \bibfield  {author} {\bibinfo {author} {\bibfnamefont {W.}~\bibnamefont
  {Cai}}, \bibinfo {author} {\bibfnamefont {L.}~\bibnamefont {Chen}}, \bibinfo
  {author} {\bibfnamefont {F.}~\bibnamefont {Ghanbarnejad}}, \ and\ \bibinfo
  {author} {\bibfnamefont {P.}~\bibnamefont {Grassberger}},\ }\href {\doibase
  10.1038/nphys3457} {\bibfield  {journal} {\bibinfo  {journal} {Nature
  physics}\ }\textbf {\bibinfo {volume} {11}},\ \bibinfo {pages} {936}
  (\bibinfo {year} {2015})}\BibitemShut {NoStop}%
\bibitem [{\citenamefont {Grassberger}\ \emph {et~al.}(2016)\citenamefont
  {Grassberger}, \citenamefont {Chen}, \citenamefont {Ghanbarnejad},\ and\
  \citenamefont {Cai}}]{Chen2016}%
  \BibitemOpen
  \bibfield  {author} {\bibinfo {author} {\bibfnamefont {P.}~\bibnamefont
  {Grassberger}}, \bibinfo {author} {\bibfnamefont {L.}~\bibnamefont {Chen}},
  \bibinfo {author} {\bibfnamefont {F.}~\bibnamefont {Ghanbarnejad}}, \ and\
  \bibinfo {author} {\bibfnamefont {W.}~\bibnamefont {Cai}},\ }\href {\doibase
  10.1103/PhysRevE.93.042316} {\bibfield  {journal} {\bibinfo  {journal} {Phys.
  Rev. E}\ }\textbf {\bibinfo {volume} {93}},\ \bibinfo {pages} {042316}
  (\bibinfo {year} {2016})}\BibitemShut {NoStop}%
\bibitem [{\citenamefont {Cui}\ \emph {et~al.}(2017)\citenamefont {Cui},
  \citenamefont {Colaiori},\ and\ \citenamefont {Castellano}}]{Cui2017}%
  \BibitemOpen
  \bibfield  {author} {\bibinfo {author} {\bibfnamefont {P.-B.}\ \bibnamefont
  {Cui}}, \bibinfo {author} {\bibfnamefont {F.}~\bibnamefont {Colaiori}}, \
  and\ \bibinfo {author} {\bibfnamefont {C.}~\bibnamefont {Castellano}},\
  }\href {\doibase 10.1103/PhysRevE.96.022301} {\bibfield  {journal} {\bibinfo
  {journal} {Phys. Rev. E}\ }\textbf {\bibinfo {volume} {96}},\ \bibinfo
  {pages} {022301} (\bibinfo {year} {2017})}\BibitemShut {NoStop}%
\bibitem [{\citenamefont {Goltsev}\ \emph {et~al.}(2006)\citenamefont
  {Goltsev}, \citenamefont {Dorogovtsev},\ and\ \citenamefont
  {Mendes}}]{Goltsev2006}%
  \BibitemOpen
  \bibfield  {author} {\bibinfo {author} {\bibfnamefont {A.~V.}\ \bibnamefont
  {Goltsev}}, \bibinfo {author} {\bibfnamefont {S.~N.}\ \bibnamefont
  {Dorogovtsev}}, \ and\ \bibinfo {author} {\bibfnamefont {J.~F.~F.}\
  \bibnamefont {Mendes}},\ }\href {\doibase 10.1103/PhysRevE.73.056101}
  {\bibfield  {journal} {\bibinfo  {journal} {Phys. Rev. E}\ }\textbf {\bibinfo
  {volume} {73}},\ \bibinfo {pages} {056101} (\bibinfo {year}
  {2006})}\BibitemShut {NoStop}%
\bibitem [{\citenamefont {Parisi}\ and\ \citenamefont
  {Rizzo}(2008)}]{Parisi2008}%
  \BibitemOpen
  \bibfield  {author} {\bibinfo {author} {\bibfnamefont {G.}~\bibnamefont
  {Parisi}}\ and\ \bibinfo {author} {\bibfnamefont {T.}~\bibnamefont {Rizzo}},\
  }\href {\doibase 10.1103/PhysRevE.78.022101} {\bibfield  {journal} {\bibinfo
  {journal} {Phys. Rev. E}\ }\textbf {\bibinfo {volume} {78}},\ \bibinfo
  {pages} {022101} (\bibinfo {year} {2008})}\BibitemShut {NoStop}%
\bibitem [{\citenamefont {Lee}\ \emph {et~al.}(2017)\citenamefont {Lee},
  \citenamefont {Choi}, \citenamefont {Kert{\'e}sz},\ and\ \citenamefont
  {Kahng}}]{Lee2017}%
  \BibitemOpen
  \bibfield  {author} {\bibinfo {author} {\bibfnamefont {D.}~\bibnamefont
  {Lee}}, \bibinfo {author} {\bibfnamefont {W.}~\bibnamefont {Choi}}, \bibinfo
  {author} {\bibfnamefont {J.}~\bibnamefont {Kert{\'e}sz}}, \ and\ \bibinfo
  {author} {\bibfnamefont {B.}~\bibnamefont {Kahng}},\ }\href@noop {}
  {\bibfield  {journal} {\bibinfo  {journal} {Scientific Reports}\ }\textbf
  {\bibinfo {volume} {7}},\ \bibinfo {pages} {5723} (\bibinfo {year}
  {2017})}\BibitemShut {NoStop}%
\bibitem [{\citenamefont {Chen}\ \emph {et~al.}(2017)\citenamefont {Chen},
  \citenamefont {Ghanbarnejad},\ and\ \citenamefont {Brockmann}}]{Chen2017}%
  \BibitemOpen
  \bibfield  {author} {\bibinfo {author} {\bibfnamefont {L.}~\bibnamefont
  {Chen}}, \bibinfo {author} {\bibfnamefont {F.}~\bibnamefont {Ghanbarnejad}},
  \ and\ \bibinfo {author} {\bibfnamefont {D.}~\bibnamefont {Brockmann}},\
  }\href {http://stacks.iop.org/1367-2630/19/i=10/a=103041} {\bibfield
  {journal} {\bibinfo  {journal} {New Journal of Physics}\ }\textbf {\bibinfo
  {volume} {19}},\ \bibinfo {pages} {103041} (\bibinfo {year}
  {2017})}\BibitemShut {NoStop}%
\bibitem [{\citenamefont {Sanz}\ \emph {et~al.}(2014)\citenamefont {Sanz},
  \citenamefont {Xia}, \citenamefont {Meloni},\ and\ \citenamefont
  {Moreno}}]{Sanz2014}%
  \BibitemOpen
  \bibfield  {author} {\bibinfo {author} {\bibfnamefont {J.}~\bibnamefont
  {Sanz}}, \bibinfo {author} {\bibfnamefont {C.-Y.}\ \bibnamefont {Xia}},
  \bibinfo {author} {\bibfnamefont {S.}~\bibnamefont {Meloni}}, \ and\ \bibinfo
  {author} {\bibfnamefont {Y.}~\bibnamefont {Moreno}},\ }\href {\doibase
  10.1103/PhysRevX.4.041005} {\bibfield  {journal} {\bibinfo  {journal} {Phys.
  Rev. X}\ }\textbf {\bibinfo {volume} {4}},\ \bibinfo {pages} {041005}
  (\bibinfo {year} {2014})}\BibitemShut {NoStop}%
\bibitem [{\citenamefont {H{\'e}bert-Dufresne}\ and\ \citenamefont
  {Althouse}(2015)}]{HebertDufresne2015}%
  \BibitemOpen
  \bibfield  {author} {\bibinfo {author} {\bibfnamefont {L.}~\bibnamefont
  {H{\'e}bert-Dufresne}}\ and\ \bibinfo {author} {\bibfnamefont {B.~M.}\
  \bibnamefont {Althouse}},\ }\href {\doibase 10.1073/pnas.1507820112}
  {\bibfield  {journal} {\bibinfo  {journal} {Proceedings of the National
  Academy of Sciences}\ }\textbf {\bibinfo {volume} {112}},\ \bibinfo {pages}
  {10551} (\bibinfo {year} {2015})}\BibitemShut {NoStop}%
\bibitem [{\citenamefont {Azimi-Tafreshi}(2016)}]{AzimiTafreshi2016}%
  \BibitemOpen
  \bibfield  {author} {\bibinfo {author} {\bibfnamefont {N.}~\bibnamefont
  {Azimi-Tafreshi}},\ }\href {\doibase 10.1103/PhysRevE.93.042303} {\bibfield
  {journal} {\bibinfo  {journal} {Phys. Rev. E}\ }\textbf {\bibinfo {volume}
  {93}},\ \bibinfo {pages} {042303} (\bibinfo {year} {2016})}\BibitemShut
  {NoStop}%
\bibitem [{\citenamefont {Serrano}\ and\ \citenamefont
  {Bogu{\~n}{\'a}}(2005)}]{Serrano2005}%
  \BibitemOpen
  \bibfield  {author} {\bibinfo {author} {\bibfnamefont {M.~{\'A}.}\
  \bibnamefont {Serrano}}\ and\ \bibinfo {author} {\bibfnamefont
  {M.}~\bibnamefont {Bogu{\~n}{\'a}}},\ }\href {\doibase
  10.1103/PhysRevE.72.036133} {\bibfield  {journal} {\bibinfo  {journal} {Phys.
  Rev. E}\ }\textbf {\bibinfo {volume} {72}},\ \bibinfo {pages} {036133}
  (\bibinfo {year} {2005})}\BibitemShut {NoStop}%
\bibitem [{\citenamefont {Chung}\ \emph {et~al.}(2014)\citenamefont {Chung},
  \citenamefont {Baek}, \citenamefont {Kim}, \citenamefont {Ha},\ and\
  \citenamefont {Jeong}}]{Chung2014}%
  \BibitemOpen
  \bibfield  {author} {\bibinfo {author} {\bibfnamefont {K.}~\bibnamefont
  {Chung}}, \bibinfo {author} {\bibfnamefont {Y.}~\bibnamefont {Baek}},
  \bibinfo {author} {\bibfnamefont {D.}~\bibnamefont {Kim}}, \bibinfo {author}
  {\bibfnamefont {M.}~\bibnamefont {Ha}}, \ and\ \bibinfo {author}
  {\bibfnamefont {H.}~\bibnamefont {Jeong}},\ }\href {\doibase
  10.1103/PhysRevE.89.052811} {\bibfield  {journal} {\bibinfo  {journal} {Phys.
  Rev. E}\ }\textbf {\bibinfo {volume} {89}},\ \bibinfo {pages} {052811}
  (\bibinfo {year} {2014})}\BibitemShut {NoStop}%
\bibitem [{\citenamefont {Kermack}\ and\ \citenamefont
  {McKendrick}(1927)}]{Kermac1927}%
  \BibitemOpen
  \bibfield  {author} {\bibinfo {author} {\bibfnamefont {W.~O.}\ \bibnamefont
  {Kermack}}\ and\ \bibinfo {author} {\bibfnamefont {A.~G.}\ \bibnamefont
  {McKendrick}},\ }\href@noop {} {\bibfield  {journal} {\bibinfo  {journal}
  {Proc. R. Soc. Lond. A}\ }\textbf {\bibinfo {volume} {115}},\ \bibinfo
  {pages} {700} (\bibinfo {year} {1927})}\BibitemShut {NoStop}%
\bibitem [{\citenamefont {Kitsak}\ \emph {et~al.}(2010)\citenamefont {Kitsak},
  \citenamefont {Gallos}, \citenamefont {Havlin}, \citenamefont {L~iljeros},
  \citenamefont {Muchnik}, \citenamefont {Stanley},\ and\ \citenamefont
  {Makse}}]{Kitsak2010}%
  \BibitemOpen
  \bibfield  {author} {\bibinfo {author} {\bibfnamefont {M.}~\bibnamefont
  {Kitsak}}, \bibinfo {author} {\bibfnamefont {L.~K.}\ \bibnamefont {Gallos}},
  \bibinfo {author} {\bibfnamefont {S.}~\bibnamefont {Havlin}}, \bibinfo
  {author} {\bibfnamefont {F.}~\bibnamefont {L~iljeros}}, \bibinfo {author}
  {\bibfnamefont {L.}~\bibnamefont {Muchnik}}, \bibinfo {author} {\bibfnamefont
  {H.~E.}\ \bibnamefont {Stanley}}, \ and\ \bibinfo {author} {\bibfnamefont
  {H.~A.}\ \bibnamefont {Makse}},\ }\href@noop {} {\bibfield  {journal}
  {\bibinfo  {journal} {Nature Physics}\ }\textbf {\bibinfo {volume} {6}},\
  \bibinfo {pages} {888} (\bibinfo {year} {2010})}\BibitemShut {NoStop}%
\bibitem [{\citenamefont {Lu}\ \emph {et~al.}(2016)\citenamefont {Lu},
  \citenamefont {Chen}, \citenamefont {Ren}, \citenamefont {Zhang},
  \citenamefont {Zhang},\ and\ \citenamefont {Zhou}}]{Lu2016}%
  \BibitemOpen
  \bibfield  {author} {\bibinfo {author} {\bibfnamefont {L.}~\bibnamefont
  {Lu}}, \bibinfo {author} {\bibfnamefont {D.}~\bibnamefont {Chen}}, \bibinfo
  {author} {\bibfnamefont {X.-L.}\ \bibnamefont {Ren}}, \bibinfo {author}
  {\bibfnamefont {Q.-M.}\ \bibnamefont {Zhang}}, \bibinfo {author}
  {\bibfnamefont {Y.-C.}\ \bibnamefont {Zhang}}, \ and\ \bibinfo {author}
  {\bibfnamefont {T.}~\bibnamefont {Zhou}},\ }\href {\doibase
  https://doi.org/10.1016/j.physrep.2016.06.007} {\bibfield  {journal}
  {\bibinfo  {journal} {Physics Reports}\ }\textbf {\bibinfo {volume} {650}},\
  \bibinfo {pages} {1 } (\bibinfo {year} {2016})},\ \bibinfo {note} {vital
  nodes identification in complex networks}\BibitemShut {NoStop}%
\bibitem [{\citenamefont {Grassberger}(1983)}]{Grassberger1983}%
  \BibitemOpen
  \bibfield  {author} {\bibinfo {author} {\bibfnamefont {P.}~\bibnamefont
  {Grassberger}},\ }\href {\doibase 10.1016/0025-5564(82)90036-0} {\bibfield
  {journal} {\bibinfo  {journal} {Mathematical Biosciences}\ }\textbf {\bibinfo
  {volume} {63}},\ \bibinfo {pages} {157} (\bibinfo {year} {1983})}\BibitemShut
  {NoStop}%
\bibitem [{\citenamefont {Newman}(2005{\natexlab{b}})}]{Newman2002}%
  \BibitemOpen
  \bibfield  {author} {\bibinfo {author} {\bibfnamefont {M.~E.~J.}\
  \bibnamefont {Newman}},\ }\enquote {\bibinfo {title} {Random graphs as models
  of networks},}\ in\ \href {\doibase 10.1002/3527602755.ch2} {\emph {\bibinfo
  {booktitle} {Handbook of Graphs and Networks}}}\ (\bibinfo  {publisher}
  {Wiley-VCH Verlag GmbH \& Co. KGaA},\ \bibinfo {year} {2005})\ pp.\ \bibinfo
  {pages} {35--68}\BibitemShut {NoStop}%
\bibitem [{\citenamefont {Karrer}\ \emph {et~al.}(2014)\citenamefont {Karrer},
  \citenamefont {Newman},\ and\ \citenamefont {Zdeborov\'a}}]{Karrer2014}%
  \BibitemOpen
  \bibfield  {author} {\bibinfo {author} {\bibfnamefont {B.}~\bibnamefont
  {Karrer}}, \bibinfo {author} {\bibfnamefont {M.~E.~J.}\ \bibnamefont
  {Newman}}, \ and\ \bibinfo {author} {\bibfnamefont {L.}~\bibnamefont
  {Zdeborov\'a}},\ }\href {\doibase 10.1103/PhysRevLett.113.208702} {\bibfield
  {journal} {\bibinfo  {journal} {Phys. Rev. Lett.}\ }\textbf {\bibinfo
  {volume} {113}},\ \bibinfo {pages} {208702} (\bibinfo {year}
  {2014})}\BibitemShut {NoStop}%
\bibitem [{\citenamefont {Martin}\ \emph {et~al.}(2014)\citenamefont {Martin},
  \citenamefont {Zhang},\ and\ \citenamefont {Newman}}]{Martin2014}%
  \BibitemOpen
  \bibfield  {author} {\bibinfo {author} {\bibfnamefont {T.}~\bibnamefont
  {Martin}}, \bibinfo {author} {\bibfnamefont {X.}~\bibnamefont {Zhang}}, \
  and\ \bibinfo {author} {\bibfnamefont {M.~E.~J.}\ \bibnamefont {Newman}},\
  }\href {\doibase 10.1103/PhysRevE.90.052808} {\bibfield  {journal} {\bibinfo
  {journal} {Phys. Rev. E}\ }\textbf {\bibinfo {volume} {90}},\ \bibinfo
  {pages} {052808} (\bibinfo {year} {2014})}\BibitemShut {NoStop}%
\bibitem [{\citenamefont {Radicchi}\ and\ \citenamefont
  {Castellano}(2016)}]{Radicchi2016}%
  \BibitemOpen
  \bibfield  {author} {\bibinfo {author} {\bibfnamefont {F.}~\bibnamefont
  {Radicchi}}\ and\ \bibinfo {author} {\bibfnamefont {C.}~\bibnamefont
  {Castellano}},\ }\href {\doibase 10.1103/PhysRevE.93.062314} {\bibfield
  {journal} {\bibinfo  {journal} {Phys. Rev. E}\ }\textbf {\bibinfo {volume}
  {93}},\ \bibinfo {pages} {062314} (\bibinfo {year} {2016})}\BibitemShut
  {NoStop}%
\bibitem [{\citenamefont {Min}(2018)}]{Min2018}%
  \BibitemOpen
  \bibfield  {author} {\bibinfo {author} {\bibfnamefont {B.}~\bibnamefont
  {Min}},\ }\href {\doibase 10.1140/epjb/e2017-80597-1} {\bibfield  {journal}
  {\bibinfo  {journal} {The European Physical Journal B}\ }\textbf {\bibinfo
  {volume} {91}},\ \bibinfo {pages} {18} (\bibinfo {year} {2018})}\BibitemShut
  {NoStop}%
\end{thebibliography}%
\clearpage

\end{document}